%% file: distribution_of_energy_revised.tex
\documentclass[twocolumn,twocolappendix]{aastex631}

\usepackage{amsmath}
\usepackage{float}

\usepackage{gensymb}
\begin{document}

\title{Distributions of energy, luminosity, duration, and waiting times of gamma-ray burst pulses with known redshift detected by \textit{Fermi}/GBM}

\input{autors}

\collaboration{20}{(AAS Journals Data Editors)}

\begin{abstract}
Discovered more than 50 years ago, gamma-ray burst (GRB) prompt emission remains the most puzzling aspect of GRB physics. Its complex and irregular nature should reveal how newborn GRB engines release their energy. In this respect, the possibility that GRB engines could operate as self-organized critical (SOC) systems has been put forward. Here, we present the energy, luminosity, waiting time, and duration distributions of individual pulses of GRBs with known redshift detected by the \textit{Fermi} Gamma-ray Burst Monitor (GBM). This is the first study of this kind in which selection effects are accounted for.
The compatibility of our results with the framework of SOC theory is discussed. We found evidence for an intrinsic break in the power-law models that describe the energy and the luminosity distributions.
\end{abstract}


\section{Introduction}
\label{sec:intro}

Gamma--ray prompt emission is the first electromagnetic radiation observed from a gamma--ray burst (GRB) source. It is caused by at least two different types of catastrophic events: (i) the merger of a binary system of compact objects \citep{Eichler89,Paczynski91,Narayan92,LIGO-Fermi17}; (ii) the core collapse of fast rotating, hydrogen-stripped massive stars (dubbed as ``collapsars'', \citealt{Woosley93,Paczynski98,MacFadyen99,Yoon05}). Most events of the first (second) class are associated with short (long) GRBs (SGRBs/LGRBs), with some notable exceptions: short with extended emission GRBs \citep{Norris06,Gehrels06}, short GRBs coming from collapsars \citep{Rossi22a,Ahumada21}, and long GRBs coming from mergers \citep{Rastinejad22,Troja22,Gompertz23,Yang23}. In the light of the emerging complexity, nowadays families (i) and (ii) are often referred to as merger and collapsar GRBs, or, equivalently, as Type-I and Type-II GRBs, respectively \citep{Zhang06_nat}.

The nature of the central engine that powers a GRB is still debated: either a hyper-accreting stellar mass black hole (BH; \citealt{Lei13,Janiuk17,Sado19}) surrounded by a thick accretion disk, or a fast-rotating, highly magnetized neutron star (NS; known as a ``millisecond magnetar'',  \citealt{Usov92,Wheeler00,Thompson04,Metzger11}).
In the case of a hyper-accreting BH, the jet can be produced either by a neutrino-dominated accretion flow (NDAF) in which neutrinos tap the thermal energy of the accretion disk via neutrino-antineutrino annihilation, launching a thermally-dominated fireball \citep{Popham99,DiMatteo02}, or through the Blandford-Znajek effect (BZ, \citealt{BlandfordZnajek77}), which converts the BH spinning energy into a Poynting-flux dominated outflow.
Also in the case of a ms-magnetar, the rotational energy could be the source of energy \citep{Duncan92,Metzger11}. Magneto-hydrodynamic (MHD) instabilities can also be considered to launch the jet, possibly for both engines~\citep{Bromberg16}.

GRB temporal profiles exhibit a remarkable variety in intensity, duration, number of pulses, and variability in general. Most profiles can be seen as a superposition of fast-rise exponential decay pulses (FRED, \citealt{FishmanMeegan95}).
Prompt emission is often characterized by phases of intense activity separated by periods of prolonged inactivity (quiescent times). 

The study of the waiting times (WTs) indicates that GRB central engine could emit pulses  according to a non-stationary Poisson process \citep{Guidorzi15b}. The diversity of all observed bursts could be understood as different realizations of a common stochastic process.
Should this be the case, an open question is whether this process can be characterized in a more detailed way, so as to reveal the dynamics that rules the energy release.
If positive, this would strongly constrain the way GRB engines work and, ultimately, help identify their nature and the powering mechanism(s).
 
Complex systems with many interacting elements, presenting erratic on-off intermittency (such as earthquakes, neuronal activity, stock market, evolution of species) are interpreted in the self-organized criticality (SOC) framework \citep{Sornette89,DeArcangelis12,Bartolozzi05,Bak93}. 
Invoked to explain the ubiquitousness of $1/f$ noise spectra, SOC was originally applied to describe sand pile avalanches \citep{Bak87,Bak88}.
A SOC system can be seen as an out-of-equilibrium nonlinear dynamical system in which energy steadily brought by an external source drives the system towards a critical point, at which the energy is liberated  through scale-free avalanches \citep{Aschwanden14b}.

One of the signatures of a SOC lies in its scale invariance: the released energy, luminosity (or peak flux, depending on the context), duration, and waiting time of individual avalanches are power-law (PL) distributed, with a set of relations between the different PL indices (see Section~\ref{sec:disc}).

SOC was pointed out to explain a wide range of astrophysical phenomena, such as solar flares, lunar craters, Saturn rings, auroral emission from Earth's magnetosphere, pulsar glitches, blazars, stellar flares (see \citealt{Aschwanden18} for a review; \citealt{AschwandenGuedel21}). One of the main accomplishments of SOC theory is the successful and exhaustive description of solar flares \citep{Lu91, Charbonneau01,Aschwanden08a, Aschwanden08b,Aschwanden10, WangDai13}.
A body of evidence for SOC behavior was also found in the case of soft gamma-ray repeaters or magnetars (SGRs; \citealt{Nakazato14}), which are also GRB engine candidates.
It is worth noting that SOC behavior in the case of a BH surrounded by an accretion disk was also investigated  as well as for the Galactic Centre Sgr~A* X-ray flaring activity \citep{Li15_SOC,Yuan18}.

Do GRB engines work as SOC systems? Cellular Automata (CA) simulations of 1D and 2D magnetized outflow demonstrated that such systems could reach a SOC state  \citep{Harko15,Danila15}. This is consistent with the picture of prompt emission being the result of fast magnetic reconnection events occurring in a Poynting outflow powered by a highly magnetized NS. Such CA simulations were also performed in the case of a hyper-accreting BH \citep{Mineshige94}, another leading candidate as GRB progenitor. Thus, the SOC interpretation could actually be consistent with radically different theoretical pictures of prompt emission.
In the case of a hyper-accreting BH, SOC avalanches could be triggered either by thermal instability of the accretion disk \citep{Janiuk07,Janiuk10,Taylor11} or by magnetic instabilities (as the kink mode instability~\citealt{Bromberg19,Giannios06b}, or the tearing mode instability~\citealt{DelZanna16,Yang19b}), while only magnetic instabilities are expected in the ms-magnetar scenario.

Evidence for SOC in GRB X-ray flares was suggested by \citet{WangDai13}, whose results could be understood as a 1D SOC system with normal diffusion. Multi-pulsed LGRBs \citep{Lyu20}, SGRBs \citep{Li23}, and LGRB precursors \citep{Li23b} seem roughly compatible with a 3D one. In these works, GRB properties are PL distributed, even if the agreement with SOC theory is not always compelling (see Table~\ref{table:comparison_results}). Moreover, observables such as fluence or peak flux of pulses are often considered, which might not necessarily reflect an intrinsic property of GRB engines, but strongly depend, among other things, on redshift. 

In this paper we used a sample of Type-II GRBs detected by the \textit{Fermi} Gamma-ray Burst Monitor (GBM; \citealt{Meegan09}) with known (spectroscopic) redshift to compute the distributions of the isotropic-equivalent released energy, of the luminosity, of the duration, and of the waiting time of the individual pulses that compose GRB light curves. We performed a thorough estimation of the selection effects, splitting our sample into redshift bins and estimating the detection efficiency separately for each redshift bin.
The redshifts of our data sample were determined through spectroscopic observations of either the optical afterglow or the host galaxy.

Section~\ref{sec:data} describes the GRB samples and the data analysis, results are reported in Section~\ref{sec:results} and discussed in Section~\ref{sec:disc}, with conclusions in Section~\ref{sec:conc}. In Appendix~\ref{sec:time-resolved} the impact of performing a time-averaged analysis instead of a time-resolved one is discussed. Hereafter, we used the flat-$\Lambda$CDM cosmology model with the latest cosmological parameters values $H_0 = 67.66~\rm{km~Mpc^{-1}~s^{-1}}$ and $\Omega_0 = 0.31$ \citep{cosmoPlanck20}.

\section{Data analysis}
\label{sec:data}

\subsection{Sample selection}
\label{sec:sample}

The GBM consists of 12 NaI scintillators sensitive in the ($\rm{8-1000~keV}$) energy range and 2 BGO scintillators, sensitive to higher energies ($\rm{150~keV-30~MeV}$), overlapping at low energies with NaI detectors. 

We first selected all the GRBs detected by the GBM from August 2008 to July 2022, with redshift measured spectroscopically, except for a few constraining photometric estimates.\footnote{They are 120922A, 151111A, and 200829A with respectively $z = 3.1 \pm 0.2$, $z = 3.5 \pm 0.3$, and $z = 1.25 \pm 0.2$.}
We excluded 130427A and 221009A from our analysis since the main burst saturates all NaI detectors of the GBM \citep{Ackermann14,Lesage23}.

To ensure that the population of GRB progenitors is as homogeneous as possible, we selected only Type-II GRBs. To this aim, we first considered events with $T_{90} \geq 2$~s \citep{Kouveliotou93}. 
We used $T_{90}$, after excluding the LGRBs that have been reported as credible Type-I candidates. The boundary value of $T_{90}=2$~s was originally derived from the catalog of the Burst And Transient Source Experiment (BATSE; \citealt{Paciesas99}): this threshold depends on the energy passband of the instrument and, as such, is consequently different for the BATSE~($\sim 2.4$ s), the Neil Gehrels Swift Observatory BAT (\citealt{Gehrels04}; $0.8$ s, see~\citealt{Bromberg13}) and the GBM ($\sim 4.2$ s, see~\citealt{vonKienlin20}). We therefore assumed the more conservative threshold of $T_{90}>4.2$~s, as it is more suitable for the GBM. $T_{90}$ values were taken from the GBM  catalog~\citep{vonKienlin20}.

Despite its $T_{90}=2.6 \pm 0.6$~s, we included 141004A because two independent telescopes (the Gran Telescopio Canarias and the Reionization and Transients Infrared Camera) detected a possible brightening in the optical counterpart, that was interpreted as an emerging supernova~\citep{Littlejohns14c,Schulze14b}. Same for 200826A with $T_{90}=1.14 \pm 0.13$~s, for which robust evidence for a collapsar origin was found \citep{Ahumada21,Zhang21a,Rossi22a}. 
Among the long-lasting GRBs we excluded 211211A, for which compelling evidence for a Type-I burst was reported despite its $T_{90}$ of 34~s and a multi-peaked light curve that deceptively looks like that of a typical Type-II GRB \citep{Rastinejad22,Troja22,Gompertz23,Yang22}.
We also excluded two GRBs (090510 and 161129A) among the so-called short with extended-emission GRBs (hereafter SEE GRBs; ~\citealt{Norris06}), that are likely Type-I GRBs in spite of their long duration due to a spectrally soft, long-lasting tail.
As a result, we were finally left with a sample of 142 Type-II GRBs.
The sample is described in Table~\ref{tab:sample}. 

\begin{table*}
\caption{\hspace{0.0cm}The first 5 GRBs of the final sample. This table is available in its entirety in machine-readable form.}
\label{tab:sample}
\begin{tabular}{lccrrccc}
\hline
 GRB & Trigger name  & Trigger time (UT) & $T_{90}(\rm{s})$ & $T_{90,\rm{err}}(\rm{s})$ & $z$ & $z_{\rm{ref}}^{\rm (a)}$\\
\hline
080804   &  bn080804972  & 23:20:14.879  &  $24.704$   &   $1.46$   &   $2.2045$ & (1) \\
080810   &  bn080810549  & 13:10:12.581  &  $75.201$   &   $3.638$  &   $3.35$   & (2) \\ 
080905B  &  bn080905705  & 16:55:46.843  &  $105.984$  &   $6.802$  &   $2.374$  & (3)\\ 
080916A  &  bn080916406  & 09:45:18.938  &  $46.337$   &   $7.173$  &   $0.689$  & (4) \\ 
080928   &  bn080928628  & 15:04:56.048  &  $14.336$   &   $4.007$  &   $1.692$  & (5)\\ 
\hline
\end{tabular}
\begin{list}{}{}
\item[$^{\rm (a)}$] (1)~\citet{Thoene08}, (2)~\citet{Prochaska08}, (3)\citet{Rowlinson10b}, (4)~\citet{Fynbo08}, (5)~\citet{Vreeswijk08b},
(6)~\citet{Davanzo08},
(7)~\citet{Kruehler15},
(8)~\citet{Berger08},
(9)~\citet{Kruehler15},
(10)~\citet{Cucchiara08},
(11)~\citet{DeUgarte09},
(12)~\citet{Kruehler15},
(13)~\citet{Chornock09},
(14)~\citet{Cenko09},
(15)~\citet{Salvaterra09},
(16)~\citet{Wiersema09},
(17)~\citet{DeUgarte09b},
(18)~\citet{Thoene09},
(19)~\citet{Cano11b},
(20)~\citet{Cucchiara09},
(21)~\citet{Fynbo09b},
(22)~\citet{DElia10},
(23)~\citet{Cucchiara09b},
(24)~\citet{Xu09},
(25)~\citet{Cucchiara09c},
(26)~\citet{Thoene09b},
(27)~\citet{Wiersema09b},
(28)~\citet{Cucchiara10b},
(29)~\citet{Kruehler13b},
(30)~\citet{Thoene10},
(31)~\citet{Kruehler13c},
(32)~\citet{Flores10},
(33)~\citet{OMeara10},
(34)~\citet{Tanvir10},
(35)~\citet{Chornock11},
(36)~\citet{DeUgarte11},
(37)~\citet{Chornock11b},
(38)~\citet{Sparre11},
(39)~\citet{Milne11},
(40)~\citet{Tanvir11},
(41)~\citet{DAvanzo11},
(42)~\citet{Chornock11c},
(43)~\citet{Klose19},
(44)~\citet{Malesani13},
(45)~\citet{Cucchiara12},
(46)~\citet{Tello12B},
(47)~\citet{DeUgarte13},
(48)~\citet{Tanvir12},
(49)~\citet{Xu12},
(50)~\citet{Tanvir12b},
(51)~\citet{Thoene12},
(52)~\citet{SanchezRamirez12},
(53)~\citet{Hartoog12},
(54)~\citet{Knust12},
(55)~\citet{Tanvir12c},
(56)~\citet{Perley12},
(57)~\citet{Cucchiara13},
(58)~\citet{DeUgarte13b},
(59)~\citet{Levan13},
(60)~\citet{Jeong14},
(61)~\citet{Smette13},
(62)~\citet{Tanvir13},
(63)~\citet{Singer13},
(64)~\citet{Sudilovsky13},
(65)~\citet{Selsing19},
(66)~\citet{Xu13},
(67)~\citet{DeUgarte13},
(68)~\citet{Malesani14},
(69)~\citet{Schulze14c},
(70)~\citet{Jeong14b},
(71)~\citet{Tanvir14},
(72)~\citet{Fynbo14},
(73)~\citet{Wiersema14},
(74)~\citet{DeUgarte14},
(75)~\citet{Singer15},
(76)~\citet{Kasliwal14},
(77)~\citet{Bhalerao14},
(78)~\citet{CastroTirado14b},
(79)~\citet{DeUgarte14b},
(80)~\citet{Gorosabel14},
(81)~\citet{Castro-Tirado14},
(82)~\citet{Xu14},
(83)~\citet{DeUgarte14d},
(84)~\citet{Gorosabel14b},
(85)~\citet{DeUgarte15},
(86)~\citet{DeUgarte15b},
(87)~\citet{Pugliese15},
(88)~\citet{DeUgarte15c},
(89)~\citet{Tanvir15},
(90)~\citet{DElia15b},
(91)~\citet{Perley15},
(92)~\citet{Bolmer15},
(93)~\citet{Tanvir16b},
(94)~\citet{Castro-Tirado16},
(95)~\citet{Xu16},
(96)~\citet{Castro-Tirado16b},
(97)~\citet{Xu16b},
(98)~\citet{Selsing19},
(99)~\citet{Castro-Tirado16c},
(100)~\citet{Malesani16b},
(101)~\citet{Cano16b},
(102)~\citet{Xu17},
(103)~\citet{Kruehler17},
(104)~\citet{DeUgarte17},
(105)~\citet{DeUgarte17b},
(106)~\citet{DeUgarte17c},
(107)~\citet{DeUgarte17d},
(108)~\citet{Melandri19a},
(109)~\citet{DeUgarte17e},
(110)~\citet{Tanvir18},
(111)~\citet{Sbarufatti18},
(112)~\citet{Izzo18b},
(113)~\citet{Izzo19b},
(114)~\citet{Vreeswijk18},
(115)~\citet{Rossi18},
(116)~\citet{Vielfaure18},
(117)~\citet{Fynbo18},
(118)~\citet{Castro-Tirado19},
(119)~\citet{Perley19},
(120)~\citet{Rossi19},
(121)~\citet{Valeev19},
(122)~\citet{Malesani19},
(123)~\citet{Yao21},
(124)~\citet{DeUgarte21},
(125)~\citet{Rossi22a},
(126)~\citet{Oates20},
(127)~\citet{Kann20b},
(128)~\citet{Kann20c},
(129)~\citet{Vielfaure20},
(130)~\citet{Vielfaure20b},
(131)~\citet{Xu21},
(132)~\citet{DeUgarte21},
(133)~\citet{Zhu21},
(134)~\citet{DeUgarte21b},
(135)~\citet{Thoene21},
(136)~\citet{Kann21},
(137)~\citet{Pozanenko21},
(138)~\citet{Fynbo22},
(139)~\citet{Castro-Tirado22},
(140)~\citet{Fynbo22b},
(141)~\citet{Saccardi22},
(142)~\citet{Izzo22b}
\end{list}
\end{table*}

\subsection{Light curve extraction and background interpolation}
\label{sec:bkgd}
The GBM has $12$ NaI detectors oriented so as to provide a nearly uniform coverage of the sky. Only a fraction of the $12$ detectors have a good view of any given GRB, typically the ones with a small angle between the normal of the detector and the direction of the GRB. To increase the signal-to-noise ratio (S/N), one conveniently sums up the most illuminated detectors' light curves. Summing too many detectors would mainly add noise and end up with a worse S/N. To find the optimal combination, we applied the  following strategy: 1) we preliminarily selected the detectors with a viewing angle $\theta < 60\degree$, as recommended by the GBM team \citep{Bhat16}, unless no detector fulfilled the previous condition or when these detectors had been used by the GBM team to compute the $T_{90}$ \citep{vonKienlin20}; 2) we generated all combinations of detectors and their corresponding light curves (e.g. if we selected detectors n1, n2, n3 at step 2, then we considered the different combinations n1+n2, n2+n3, n1+n3 and n1+n2+n3); 3) we took the combination with most pulse candidates (with $\rm{S/N} > 6$). In case of equality, we opted for the combination of detectors that maximizes the total S/N of all identified pulse candidates. 
For each GRB, we extracted the Time-Tagged Event (TTE) light curves for the selected NaI detectors in three different energy channels: 8--1000, 30--1000, and 40--1000~keV. TTE data typically cover from $-30$ to $300$~s with respect to the trigger time. For some extended triggers, TTE data coverage extend from $-100$ to $450$~s.
For very long GRBs ($T_{90}\gtrsim 600$~s), TTE data do not cover the full extent of the events, as in the case of the ultra-long 091024 \citep{Virgili09,Gruber11a}, so we used instead the CSPEC data\footnote{CSPEC data are continuous high
spectral resolution data with 1.024 s time resolution during bursts.}. For the ultra-long 160625B (e.g., \citealt{Zhang18e}) we used TTE data for the first part of the event ($t<400$~s) and CSPEC data for the second part ($t>400$~s).

The light curves were extracted using the GBM tools publicly available\footnote{\url{https://fermi.gsfc.nasa.gov/ssc/data/analysis/gbm/gbm_data\_tools/gdt-docs/}} \citep{GbmDataTools}. We binned these light curves with a time resolution from 16 to 1024 ms. Some GRBs had particularly short timescales that required the higher resolution of 4~ms\footnote{It was the case for 090424, 090902B, 090926, 130427A, 180720B, 190114C.} (see Section~\ref{sec:peaks} for more details on the identification of the optimal bin time for each GRB).

In order to interpolate the background in the ``on-source" time interval (where the burst is present), we chose two ``off-source" windows and modeled the background with a polynomial function with order up to 3. To check the quality of the interpolation, we computed the normalized residuals $\epsilon_i$ in the off-source windows defined as:
\begin{equation}
    \epsilon_{i} = \frac{r_{i} - b_{i}}{\sigma_{{r_{i}}}}\;,
\label{residui_bkgd}
\end{equation}
where $r_{i}$ is the count rate in the $i$-th bin, $\sigma_{r_{i}}$ is the count rate uncertainty in the $i$-th bin, given by:
\begin{equation}\label{sigma}
    \sigma_{
    r_i} = \sqrt{\frac{r_i}{\Delta t}}\;,
\end{equation}
obtained assuming a Gaussian regime in counts per bin time $\Delta t$, and $b_{i}$ is the interpolated background count rate.
An optimal background modeling is characterized by normally distributed residuals with null mean value and unity standard deviation $\mathcal{N}(0,1)$ (standardized normal distribution). 

The background-subtracted light curve was considered for the next steps in the analysis, provided that the following conditions on the mean value $\mu_\epsilon$ and standard deviation $\sigma_\epsilon$ were fulfilled: $|\mu_\epsilon| < 0.2$ and $|\sigma_\epsilon -1| < 0.2$. Otherwise, the corresponding detector was ignored.

We obtained a background-subtracted light curve independently for each detector and energy passband and then we added them through all the combinations mentioned above to obtain as many background-subtracted light curves. We selected for the following study the light curve associated with the best combination of detectors.

%---------------------------------------------
\subsection{Pulse identification}
\label{sec:peaks}
%---------------------------------------------
Pulses were identified using {\sc mepsa} \footnote{This code is registered at the ASCL with the code entry ASCL 1410.002
 and is also available at \url{https://www.fe.infn.it/u/guidorzi/new_guidorzi_files/code.html}}
\citep{Guidorzi15a}, that is a peak search algorithm designed and calibrated to find statistically significant local maxima in GRB light curves. Each pulse candidate is characterized in terms of the pulse peak time $t_0$, the pulse peak rate $A$, the S/N, and the detection timescale at which the pulse was detected with the highest S/N, hereafter denoted $\Delta t_{\rm det}$.  For the goal of the present analysis we rejected all the pulse candidates with S/N $<6$ to ensure a high purity.

Since {\sc mepsa} applies to uniformly binned light curves, we preliminarily had to determine the optimal bin time for any given GRB. On the one hand, using an unnecessarily short bin may lead to relatively numerous statistical fluxes. On the other hand, a too coarse resolution would wash out genuine temporal structures with a consequent loss of information.
We opted for the longest bin time $\Delta t$ for which all the pulses detected by {\sc mepsa} are resolved. This condition was implemented by asking $\Delta t \leq \Delta t_{{\rm det},i}/2$ for any generic $i$-th pulse candidate. As for the choice of the combination of detectors, we took the one with most pulse candidates, as explained in Section~\ref{sec:bkgd}.

We applied {\sc mepsa} to the light curves in the three energy passbands mentioned in Section~\ref{sec:sample} and with a bin time in the range 16--1024 ms, with a maximum rebin factor of 100 (see \citealt{Guidorzi15a} for details), except for the few GRBs which required a 4-ms resolution (see Section~\ref{sec:bkgd}). In harder channels, pulses tend to be narrower \citep{Fenimore95}: we exploited this property to improve our ability in identifying and separating partially overlapped pulses.

Since most photons are relatively soft due to GRB spectral shapes, neglecting the softer energy channels turns into a statistically poorer S/N. We therefore chose to merge the results obtained with {\sc mepsa} on the three different energy ranges, whenever visual inspection suggested the presence of blended pulses that were missed by {\sc mepsa} in the analysis of individual energy channels. We made sure to count just once the pulses that were identified in more than one energy band. To this aim, we adopted the following strategy: let $t_{0i}$ and $\Delta t_{{\rm det},i}$ be respectively the pulse peak time and the detection timescale of the $i$-th pulse candidate detected by {\sc mepsa} in  the 8--1000~keV band. Let $t'_{0j}$ and $\Delta t'_{{\rm det},j}$ be the analogous quantities 
of the $j$-th pulse candidate from one of the harder channels. The two pulse candidates are tagged as distinct ones if their intervals do not overlap:
\begin{equation}
\label{eq:interval}
    \Big[t_{0i} - \frac{\Delta t_{{\rm det},i}}{2} \ ;\ t_{0i} + \frac{\Delta t_{\rm{det,i}}}{2}\Big] \cap\Big[t'_{0j} - \frac{\Delta t'_{{\rm det},j}}{2}\ ;\ t'_{0j}+ \frac{\Delta t'_{{\rm det},j}}{2}\Big] = \emptyset\;.
\end{equation} 

The condition~\eqref{eq:interval} eliminates most of multiple detections of the same pulse, except for a few cases, which were corrected after visual inspection.

%---------------------------------------------
\subsection{Light curve modeling}
\label{sec:fit}
%---------------------------------------------
We modeled each identified pulse candidate with a FRED profile, which was found to describe most GRB pulses \citep{Norris96}:
\begin{equation}
\label{eq:norris}
  N(t| t_0,A,\tau_r,\xi,\nu)\, =\ A
    \begin{cases}
       \exp{\Big[-\Big(\frac{|t-t_{0}|}{\tau_{r}}\Big)^{\nu}\Big]}  & (t\leq t_0)\\

      \exp{\Big[-\Big(\frac{|t-t_{0}|}{\xi\tau_{r}}\Big)^{\nu}\Big]}  & (t>t_0)\;,
    \end{cases}
\end{equation}

where $A$ is the pulse peak rate, $t_0$ the pulse peak time, $\tau_r$ the pulse rise time, $\xi$ the decay-to-rise ratio (note that the pulse decay time is $\tau_d = \xi\tau_r$), and $\nu$ the so-called peakedness, which determines the pulse sharpness. As first noted by \citet{Norris96} in the analysis of BATSE bursts, $\xi$ mostly ranges between 2 and 3, whereas $\nu$ lies between 1 (pure exponential) and 2 (Gaussian). Finally, we investigated the impact of the choice of the pulse model by alternatively adopting the one proposed in \citet{Norris05}. In this model two parameters (two timescales) instead of three are needed to define the shape of a pulse.

We used a non-linear least square algorithm\footnote{The fit was done using {\tt scipy.optimize.curve\_fit} from the {\tt scipy} library~\citep{Virtanen20}.} to fit the light curves with a superposition of Norris pulses, as in Equation~\eqref{eq:sum_fred}:
\begin{equation}
\label{eq:sum_fred}
    f(t) = \sum_{i=1}^{N_{\rm p}} N(t| t_{0i},A_i,\tau_{ri},\xi_i,\nu_i)
\end{equation}

The number of pulses $N_{\rm p}$ was given by the results of {\sc mepsa}. We used the pulse peak times and the pulse peak rates given by {\sc mepsa} as starting values for the fit. We set some boundaries on the $5N_{\rm{p}}$ parameters (5 for each pulse). The pulse peak time of a generic $i$-th pulse was constrained to be in the interval $[t_{0i} - \Delta t_{{\rm det},i}/2\ ;\ t_{0i} + \Delta t_{{\rm det},i}/2]$. The pulse peak rate was allowed to vary within the interval $[A_i - n_2  \sigma_{A_i} ; A_i + n_2 \sigma_{A_i}]$ where $\sigma_{A_i}$ is the pulse peak rate error given by {\sc mepsa}, whereas $n_2$ can be adjusted by the user. For well separated pulses we used $n_2 = 1$. Instead, for partially overlapped pulses we allowed larger values for $n_2$. For the GRB light curves consisting of a forest of overlapped pulses, we left the pulse peak rate unconstrained. To avoid unrealistic modeling of pulses, we constrained $\xi$ and $\nu$ in the following ranges: $0.1<\xi<10$ and $0.4<\nu<4$, as suggested by the results by \citet{Norris96} in their systematic analysis of BATSE GRBs, which shared the energy passband with GBM NaI detectors.

Similarly to the validation of the background modeling (Section~\ref{sec:bkgd}), we here verified the quality of the fit by computing the corresponding normalized residuals defined as:
\begin{equation}
\label{eq:norm_res_model}
    \epsilon_i = \frac{r_i - f(t_i)}{\sigma_{r_i}}
\end{equation}

where $r_i$ is the $i$-th count rate, $\sigma_{r_i}$ its associated uncertainty, $f(t_i)$ the count rate predicted by the model of Equation~\eqref{eq:sum_fred} at time $t_i$.

\begin{figure}
 \includegraphics[width=\columnwidth]{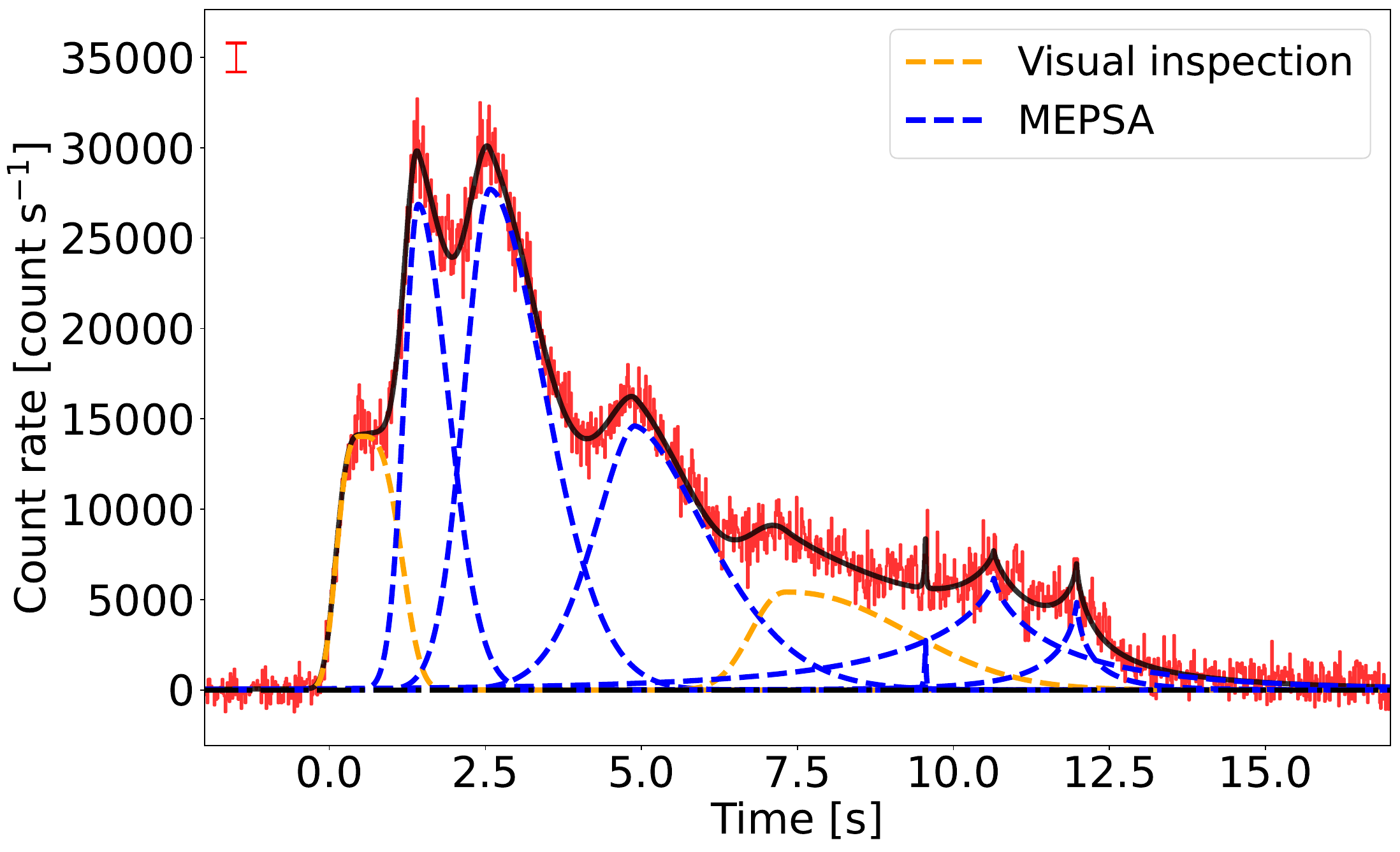}
 \caption{150314A background-subtracted light curve (red) along with the best-fit model (black). The typical error size is shown in the top left. The black horizontal line represents the interpolated background. Blue dashed curves represent the individual pulses detected by {\sc mepsa} while orange ones represent pulses added by visual inspection to improve the fit of this complex GRB time profile. }
  \label{fig:090113}
\end{figure}

We computed the mean, median, and standard deviation of $\{\epsilon_i\}$ over the same time interval that was used for the background interpolation plus the one that includes the GRB profile. Ideally, $\epsilon_i$ is distributed as a standardized Gaussian ${\cal N}(0,1)$. A non-zero mean would reveal unaccounted trends in the modeling, whereas a value of $\sigma>1$ ($\sigma<1$) would indicate under(over)-fitting. A poor modeling often turned out to be due to the presence of weak and/or blended unaccounted pulses: in such cases we applied the strategy described in Section~\ref{sec:peaks} to add pulses to the signal until we reached an acceptable solution. In some cases we had to add pulses by visual inspection, until the quality of the fit improved sufficiently. Figure~\ref{fig:090113} shows an example.
For a few challenging GRBs we could only come up with a relatively poor fit. More details are reported in Section~\ref{sec:results}.

For each GRB we obtained a set of $5 N_{\rm p}$ best-fit parameters $\boldsymbol{\mu_{\rm fit}}$ with an associated covariance matrix $\boldsymbol{\Sigma_{\rm fit}}$, which was used to estimate the uncertainties on the best-fit parameters. The uncertainties on the best fit parameters are given by the square root of the diagonal of the covariance matrix. For each GRB we randomly generated a sample of $10^{3}$ sets of parameters around the best-fit solution according to a multivariate Gaussian distribution ${\cal N}(\boldsymbol{\mu_{\rm fit}},\boldsymbol{\Sigma_{\rm fit}})$ truncated along the positive-definite parameters. 
Once the best-fit profile of Equation~\eqref{eq:norris} is known for every pulse, we computed the integral counts of the $i$-th pulse:
\begin{equation}
\label{eq:equation_C_i}
    C_i\ =\ \int_{-\infty}^{+\infty} N(t| t_{0i},A_i,\tau_{r,i},\xi_i,\nu_i) \,dt\;.
\end{equation}
The uncertainty on $C_i$ was estimated as the standard deviation of the corresponding distribution of values generated by random realizations of the set of parameters using ${\cal N}(\boldsymbol{\mu_{\rm fit}},\boldsymbol{\Sigma_{\rm fit}})$.

%-------------------------------------------------
\subsection{Isotropic-equivalent energy and luminosity}
\label{sec:spec}
%-------------------------------------------------
In the following, energy flux (erg~cm$^{-2}$~s$^{-1}$) and fluence (erg~cm$^{-2}$) are denoted with $F$ and $\Phi$, respectively.

We estimated the fluence of the $i$-th pulse as $\Phi_i = C_i\, f_{\rm tot}$, where $f_{\rm tot} = \Phi_{\rm tot}/C_{\rm tot}$ is the conversion factor obtained from the corresponding time-integrated quantities. $\Phi_{\rm tot}$ is the fluence of the whole GRB and $C_{\rm tot}$ the corresponding counts calculated on the same time interval. 

$\Phi_{\rm tot}$ was estimated by taking the best fit model and its associated spectral parameters provided by the GBM GRB catalog\footnote{The considered spectral models are PL, Band, Comptonized, and Smooth Broken Power Law. Their description is available at \url{https://fermi.gsfc.nasa.gov/ssc/data/analysis/gbm/gbm\_data\_tools/gdt-docs/api/api-spectra.html\#spectral-functions} .}. For 091024 we used instead fluence values on the whole event provided by \citet{Gruber11b}. K-corrections, needed to estimate the fluence in the common rest-frame energy passband 1--$10^4$~keV, were computed in the following way:
\begin{equation}
\label{eq:kcorrection}
    k_c = \frac{ \int_{\frac{1}{1+z}}^{\frac{10^{4}}{1+z}} EN(E) \,dE }{ \int_8^{1000} EN(E) \,dE}\;.
\end{equation}
This choice complies with what is usually adopted in the GRB literature (e.g., \citealt{Amati02}). Therefore, we computed $\Phi_{\rm tot}$ as $\Phi_{\rm tot} = k_c~\Phi_{\rm{8-1000}}$.
This procedure assumes a negligible impact of the spectral evolution throughout the GRB: in Appendix~\ref{sec:time-resolved} we show that the impact of our approximation on the resulting pulse energy distribution does not affect our conclusions significantly.

The corresponding released isotropic-equivalent energy is calculated as
\begin{equation}
    E_{{\rm iso},i}\ =\  \frac{4\pi d_{L}^{2}}{1+z}\ \Phi_i\;,
\label{eq:eiso_time_avgd}
\end{equation}
%,
where $d_L$ is the luminosity distance\footnote{The luminosity distance was computed with cosmological parameters mentioned in Section~\ref{sec:intro} using the package {\tt Planck18} of the  {\tt astropy.cosmology} library.}.
The corresponding isotropic-equivalent luminosity is computed as
\begin{equation}
    L_{{\rm iso},i}\ =\  4\pi d_{L}^{2}\ F_i\;,
\label{eq:Liso_time_avgd}
\end{equation}
where $F_i = A_i\, f_{\rm{tot}}\;$.

%%%%%%%%%%%%%%%%%%%%%%%%%%%%%%%%%%%%%%%%%%%%%%%%%%
\section{Results}
\label{sec:results}
%%%%%%%%%%%%%%%%%%%%%%%%%%%%%%%%%%%%%%%%%%%%%%%%%%
{\sc mepsa} detected 974 pulses in the 8--1000~keV band, while 13 additional pulses were detected in the harder channels.
We had to add at least one pulse by visual inspection for 45 GRBs. Overall, we added 159 undetected pulses to the 987 detected by {\sc mepsa}, so that the final sample consists of 1146 pulses, whose 14\% had to be identified through visual inspection.

We split the GRB sample in two groups, depending on the quality of the modeling and on the presence of pulses with extreme values of the parameters which define the pulse shape ($\xi$ and $\nu$). The Silver Sample\footnote{They are 080810, 090328, 090927, 091003, 091024, 100414A, 100728A, 120711A, 130528A, 131108A, 140304A, 141220A, 150403A, 160629A, 161117A, 170214A, 171010A, 180205A, 180720B, 180728A, 181020A, 220101A, and 220627A.}
 (hereafter, SS) includes both (i) GRBs with a standard deviation of the normalized residuals deviating from 1 by at least $0.1$: $|\sigma_\epsilon-1| > 0.1$, and (ii) GRBs for which the best-fit model contains at least one pulse with an extreme shape in terms of parameters $\xi$ and $\nu$ (either $\xi < 0.11$ or $\xi > 8.99$ and $\nu < 0.41$ or $\nu > 3.99$ respectively). The Golden Sample (hereafter, GS) includes all the remaining GRBs.

The GS (SS) consists of 119 (23) GRBs. In terms of pulses, the GS (SS) includes 696 (450) pulses. On average, SS GRBs have more pulses per GRB, with a median of 15 against 3 of the GS. They are also brighter and more energetic with a median fluence $\sim 4.5$ times higher ($4.4 \times 10^{-5}$ vs. $9 \times 10^{-6}$~erg~cm$^{-2}$). The isotropic-equivalent energy of SS GRBs is on average $\sim 7$ times bigger ($4.3 \times 10^{53}$ vs. $6.2 \times 10^{52}$ ~erg). The fact that, on average, the GRBs having the best signal are also the most problematic ones to model, reveals the degree of complexity of GRB time profiles as well as the limits of our approach. 

\begin{figure}
 \includegraphics[width=\columnwidth]{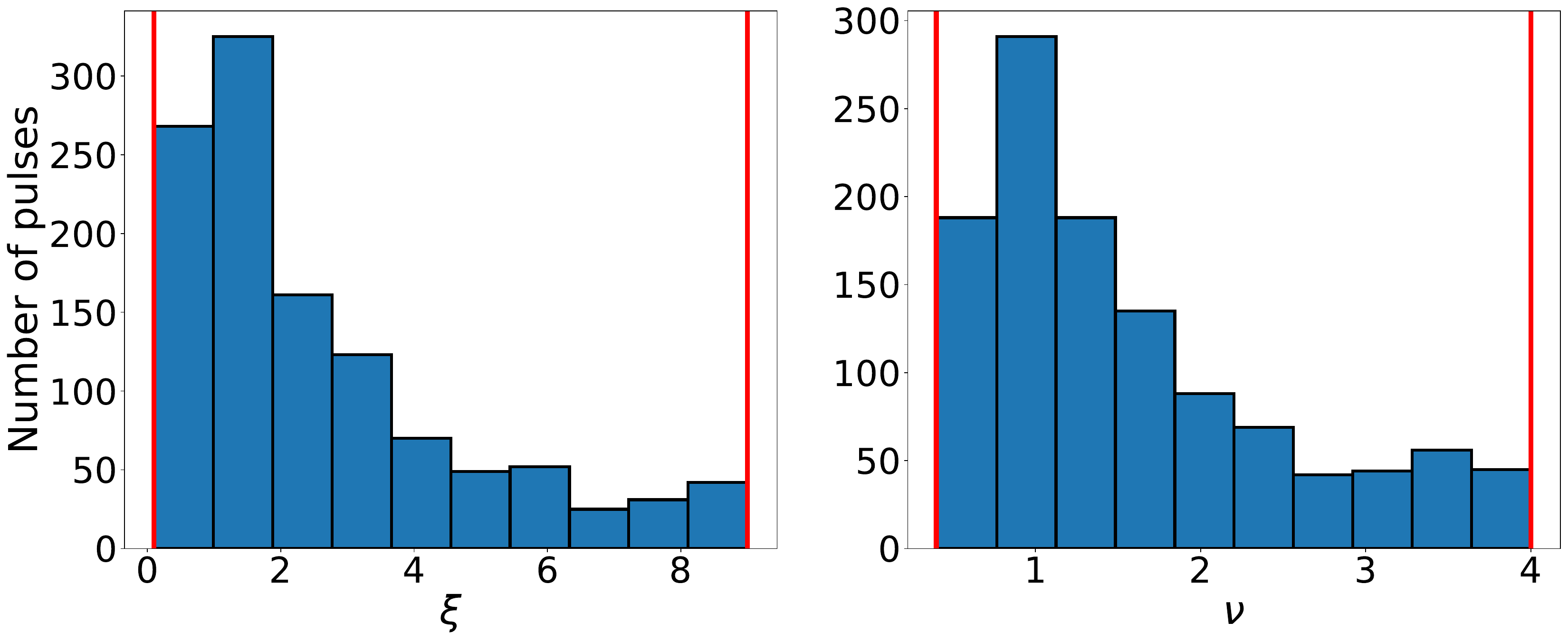}
 \caption{$\xi$ and $\nu$ distributions for the whole sample. Red vertical lines indicate the chosen parameter boundaries ($0.1<\xi<9$ and $0.4<\nu<4$).}
  \label{fig:r_nu_distrib}
\end{figure}

Our mean values of $\langle\xi\rangle$ and $\langle\nu\rangle$ for the whole sample (GS+SS) are respectively $2.6$ and $1.6$, which is broadly compatible with the corresponding distribution obtained by \citet{Norris96} for BATSE GRBs (Figure~\ref{fig:r_nu_distrib}).

%-------------------------------------------------
\subsection{Detection efficiency}
\label{sec:deteff}
%-------------------------------------------------
To account for the completeness limit of the energy and of the luminosity distributions, we estimated the detection efficiency as a function of the pulse peak rate $A$. In the case of energy, the detection efficiency is a function of the pulse counts $I$. 

In either case we estimated {\sc mepsa} efficiency to detect features on samples of simulated pulses with different shapes for given pulse peak rates/pulse counts. We carried out two complementary kinds of simulations: (i) single- and (ii) multi-pulse.

For (i), for every GRB we took its interpolated background profile and added a randomly generated FRED pulse. We finally added Poisson noise. 

For (ii), for every GRB we randomly generated and added a FRED pulse to the best-fit model of the real GRB light curve plus the modeled background curve. Hence, we ended up with as many replicates of the original GRB light curves, each of which included one additional synthetic pulse. This was done in order to mimic the more realistic situation in which a given pulse might have occurred in conjunction with the other ones that were already identified. We finally added Poisson noise.

Each simulated pulse was generated assuming for $t_r$, $\xi$, and $\nu$ the corresponding log-normal distributions that best fit the observed distributions of the same parameters in the real samples. These best-fit log-normal distributions were validated through a two-population Kolmogorov-Smirnov (KS) test.\footnote{The KS test was done using {\tt scipy.stats.ks\_2samp}.}
A random variable $X$ is log-normally distributed if $\ln({X}) \sim \mathcal{N}(\mu,\sigma^2)$. The best fit log-normal distribution parameters ($\mu$, $\sigma$) are ($-0.69$, $1.5$), ($0.66$, $0.97$), and ($0.35$, $0.62$) for  $t_r$ expressed in s, $\xi$, and $\nu$ distributions, respectively. 

In the case of energy, the simulated pulse peak rate $A$ was computed as a function of a given $I$, which is the total counts assigned to a given pulse. The rationale behind this choice is to describe the detection efficiency for pulses with a given $I$, which is the most important parameter for a pulse to be detected (against a given background rate). Since $I$ scales linearly with $A$ (see Equation~\eqref{eq:equation_C_i}), this is easily done:
\begin{equation}
I_1\  =\ \int_{-\infty}^{+\infty} N(t  ;A_1,t_0,t_r,\xi,\nu) \,dt\;, 
\end{equation}
where we set $A_1 = 1$. $A$ is then calculated as follows:
\begin{equation}
\label{eq:renorm_A}
A  = A_1\frac{I}{I_1}\;.
\end{equation}
In the case of luminosity, as pulses of different shapes are grouped by count rates, we simply assigned a given pulse peak rate to a given group of simulated pulses.

For (i), the pulse peak time was set to $ t_0 = 0~\rm{s}$ by convention, as its position is not relevant in the single-pulse case. For (ii), the pulse peak time was drawn from a uniform distribution with the constraint to obtain pulses with separability (defined as the distance between the simulated pulse and its nearest neighbor divided by its FWHM) between $1$ and $10$.

For each GRB we generated a sample of $100$ simulated FRED-like pulses with parameters $[t_0,t_r,\xi,\nu]$ randomly chosen from the corresponding best-fit log-normal distributions. 

We repeated the process for evenly spaced logarithmic values of $I$ ($A$), in the range $10^2$--$10^4$ counts ($10^2$--$10^4$~counts~s$^{-1}$). We used the same sample of FRED parameters for each value of $I$ or $A$.
This procedure allowed us to model the detection efficiency as a function of the pulse counts and of the pulse peak rate of a generic pulse.
We finally obtained the detection efficiency as a function of either $E_{\rm iso}$ or $L_{\rm iso}$, using Equations~\eqref{eq:eiso_time_avgd} and \eqref{eq:Liso_time_avgd} separately for $9$ bins in redshift\footnote{The bin edges are given by [0.078    , 0.139, 0.238, 0.400  , 0.662,
       1.082, 1.76, 2.898, 4.813, 8.100]}: these were obtained through an evenly spaced logarithmic sampling of the luminosity distance range.

The detection efficiency $\eta_z(E_{\rm iso})$ ($\eta'_z(L_{\rm iso}))$ was estimated as the fraction of simulated pulses with $E_{\rm iso}$ ($L_{\rm iso}$) (taken from the grid of simulated values) that had been identified by our procedure at the given redshift bin. Finally, $\eta_z(E_{\rm iso})$ ($\eta'_z(L_{\rm iso})$) was obtained for any value of $E_{\rm iso}$ ($L_{\rm iso}$) through interpolation.\footnote{We applied a kernel density estimate with Gaussian kernel as implemented in python class {\tt scipy.stats.gaussian\_kde}.} 

%-------------------------------------------------
\subsection{Modeling of energy and luminosity distributions}
\label{sec:modeling}
%-------------------------------------------------
Having modeled the redshift-dependent selection effects that are present in the observed $E_{\rm iso}$ and $L_{\rm iso}$ distributions, we now aim to test if/which given conjectural intrinsic differential distribution $f(E_{\rm iso})=dN/dE_{\rm iso}$ ($f'(L_{\rm iso})=dN/dL_{\rm iso}$)  may explain the results, once the selection effects are properly simulated. 
To this aim, we started with the simplest distribution, that is a PL: $f(E_{\rm iso})\propto E_{\rm iso}^{-s}$ ($f'(L_{\rm iso})\propto L_{\rm iso}^{-\kappa}$), with $s$ and $\kappa$ being the energy and luminosity PL indices. As an alternative, we considered a broken power law (BPL): $~ f(E_{\rm iso})\propto E_{\rm iso}^{-s_1}$ for $E_{\rm iso} < E_b$ and $f(E_{\rm iso}) \propto E_{\rm iso}^{-s_2}$ for $E_{\rm iso} \geq E_b$, where $s_1$ and $s_2$ are the low- and high-energy indices, respectively, and $E_b$ the break energy. We firstly tried to fix the low-energy index to zero, $s_1=0$, such that the resulting BPL is similar to a thresholded PL \citep{Aschwanden15}, although not mathematically equivalent to it. Similarly, we considered a BPL distribution for the luminosity distribution: $~ f'(L_{\rm iso})\propto L_{\rm iso}^{-\kappa_1}$ for $L_{\rm iso} < L_b$ and $f'(L_{\rm iso}) \propto L_{\rm iso}^{-\kappa_2}$ for $L_{\rm iso} \geq L_b$, where $\kappa_1$ and $\kappa_2$ are the low- and high-luminosity indices, respectively, and $L_b$ the break luminosity.  

For each redshift bin, we assigned a detection probability $p = \eta_z(E_{\rm iso})$~($p' = \eta'_z(L_{\rm iso})$) to any given value of $E_{\rm iso}$ ($L_{\rm iso}$), that was randomly generated from $f(E_{\rm iso})$ ($f'(L_{\rm iso})$). Each simulated value of $E_{\rm iso}$ ($L_{\rm iso}$) was kept, provided that a Bernoulli trial\footnote{In other words, if we let S be the random variable that equals $1$ if a given pulse with energy $E_{\rm iso}$ is detected and $0$ otherwise, $S \sim \mathcal{B}(p)$ with $p = P(S=1) = \eta_z(E_{\rm iso})$.} with the probability of success given by $\eta_z(E_{\rm iso})$ ($\eta'_z(L_{\rm iso})$) turned out to be $1$.
This process was iterated until we ended up with as many pulses as in the observed distribution for that redshift bin. Consequently, the resulting simulated distribution has the same number of events and same redshift distribution as the observed distribution.
\begin{figure*}
 \includegraphics[width=18cm]{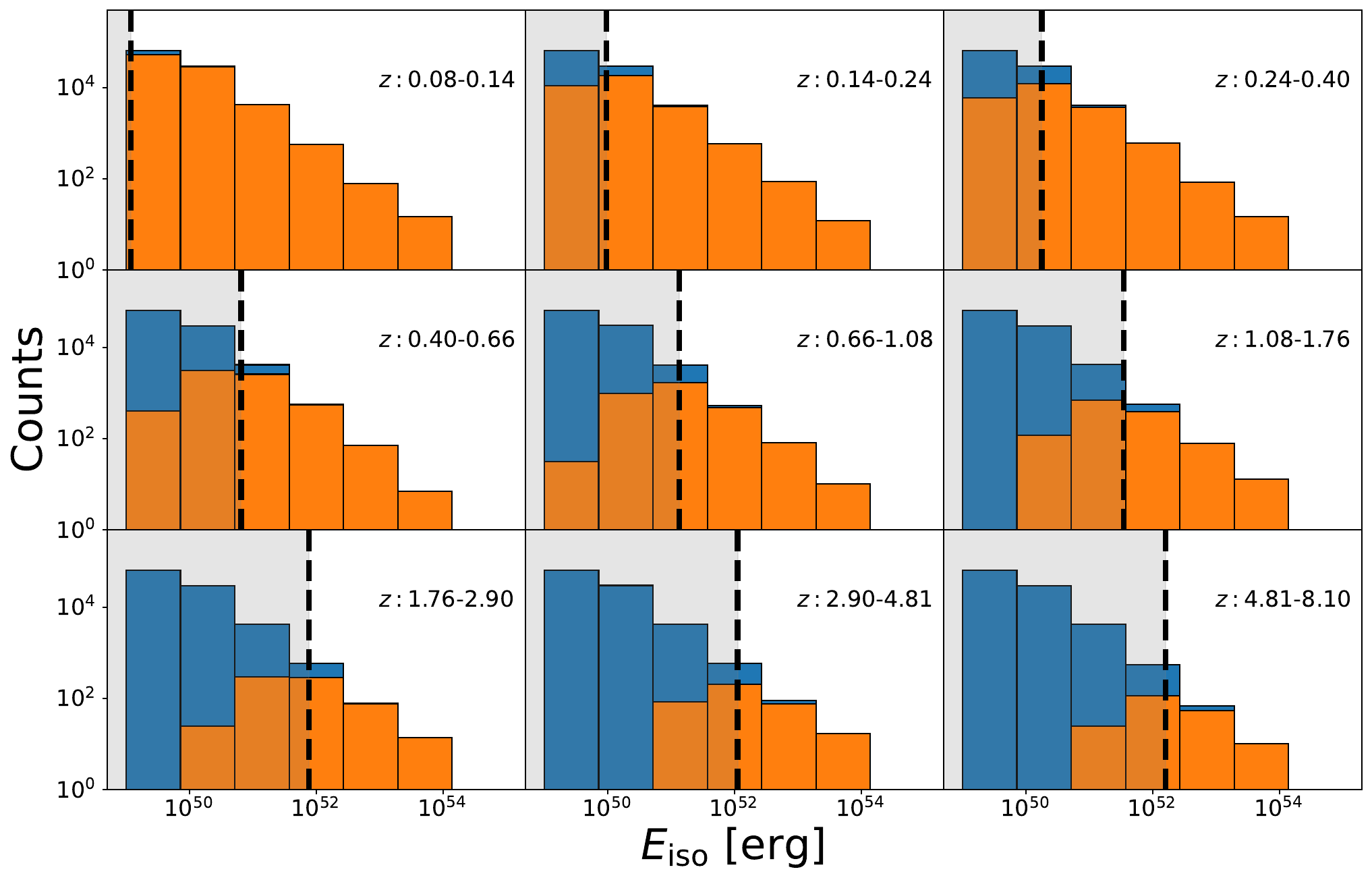}
 \caption{Simulated distributions of $E_{\rm iso}$, depending on whether selection effects due to detection efficiency are considered (orange) or not (blue). The nine panels refer to increasing redshift bins (left to right, top to bottom). The blue distribution was drawn assuming $dN/dE_{\rm iso}\propto E_{\rm iso}^{-2}$. Vertical dashed lines mark the 50\% detection efficiency: pulses with lower values of $E_{\rm iso}$, highlighted by the gray area, are therefore hampered by a low detection efficiency. For the sake of clarity, $10^{5}$ pulses were simulated in each redshift bin.}
\label{fig:selection_effects_redshift}
\end{figure*}
To appreciate the impact of the selection effects, Figure~\ref{fig:selection_effects_redshift} displays the simulated energy distribution for each of the 9 different redshift bins, along with the one predicted assuming $dN/dE_{\rm iso}\propto E_{\rm iso}^{-2}$ and the same number of GRBs per each redshift bin.

%-----------------------------------------------
\subsubsection{Validation of a putative intrinsic differential distribution}
\label{sec:validation_distrib}
%-----------------------------------------------
To assess the probability that the observed $E_{\rm iso}$ ($L_{\rm iso}$) distribution is the result of an assumed intrinsic differential distribution $f(E_{\rm iso})$ ( $f'(L_{\rm iso})$) net of selection effects, we carried out a few tests: likelihood ratio test (LRT), $\chi^{2}$, two-population KS, and Anderson-Darling\footnote{The two-sample AD test was done using {\tt scipy.stats.anderson\_ksamp}} (AD) tests.

We divided the range of $E_{\rm iso}$ ($L_{\rm iso}$) values into $M = 6$ (6) logarithmic evenly spaced bins, ensuring that each bin contained $\ge20$ pulses. Let $C_{o,i}$ ($C_{s,i}$) be the number of pulses in bin $i$ of the observed (simulated) distribution, either in the case of $E_{\rm iso}$ or $L_{\rm iso}$. Let $N_{\rm o}=\sum_{i=1}^{M} C_{o,i}$ and $N_{\rm s}=\sum_{i=1}^{M} C_{s,i}$ the total number of pulses in the observed and simulated distributions, respectively. For any simulated distribution $s$
the $\chi^{2}_s$ and the LRT$_s$ are obtained by computing the following quantities:

\begin{equation}
   \chi^{2}_s = \sum_{i=1}^{M} \frac{(C_{o,i}-C_{s,i})^{2}}{C_{o,i}+C_{s,i}}\;,
   \label{eq:chi2}
\end{equation}

\begin{equation}
   {\rm LRT}_s =  2\, \log~\prod_{i=1}^{M}\frac{ \biggl(\frac{C_{o,i}}{N_o}\biggl)^{C_{o,i}}\biggl(\frac{C_{s,i}}{N_s}\biggl)^{C_{s,i}}}{ \biggl(\frac{C_{o,i}+C_{s,i}}{N_o+N_s}\biggl)^{C_{o,i}+C_{s,i}}}\;.
   \label{eq:LRT}
\end{equation}

Both metrics were calculated for a sample of $100$ simulated distributions. We then took the median values of both. In Equation~\eqref{eq:LRT}, the numerator represents the maximum likelihood under the assumption that the two sets are independently distributed, whereas the denominator is the maximum likelihood under the alternative assumption that both sets share the parent distribution. Under the assumption that the two sets have a common distribution, LRT is $\chi^2_{M-1}$ distributed:\footnote{This is the reason why we incorporated a factor of $2$ in Equation~\eqref{eq:LRT}.} we therefore carried out a one-tail test, calculating the p-value as $P(\chi^2_{M-1}\ge {\rm LRT})$. 

In the case of the $E_{\rm iso}$ distribution, we were not able to find a satisfactory solution for the PL model according to all of the four tests: in particular, the best PL model yielded a value of $\sim 60$ for both $\chi^2$ and LRT (Equations~\eqref{eq:chi2} and \eqref{eq:LRT}). Consequently, a simple PL model is confidently rejected.

Moving to the BPL model, given that two additional parameters come into play, we initially considered two alternative approaches: (1) we fixed $s_2 = 1.5$ and let both $s_1$ and $E_b$ free to vary; (2) we fixed $s_1 = 0$ and let both $s_2$ and $E_b$ free.  This was done to make a preliminary exploration of the parameter space. Then we allowed all parameters to vary and performed Markov Chain Monte Carlo (MCMC) simulations with {\sc emcee} \citep{Foreman13} to sample the LRT around its minimum value. We took a Gaussian ball around the minimum LRT value of model (1) and performed a MCMC run with 1000 samples and 64 walkers. Then we compute the LRT on a cube of parameters around the minimum LRT value given by the MCMC run to find the minimum LRT value and the confidence intervals.
We found that the BPL that minimized the $\chi^{2}$ and the LRT was around $s_1 \sim 1$, $s_2 \sim 1.67$ and $E_b \sim 1\times 10^{52}$~erg, with $\chi_{5}^{2}/5=5.25/5$.

\begin{figure}
 \includegraphics[width=\columnwidth]{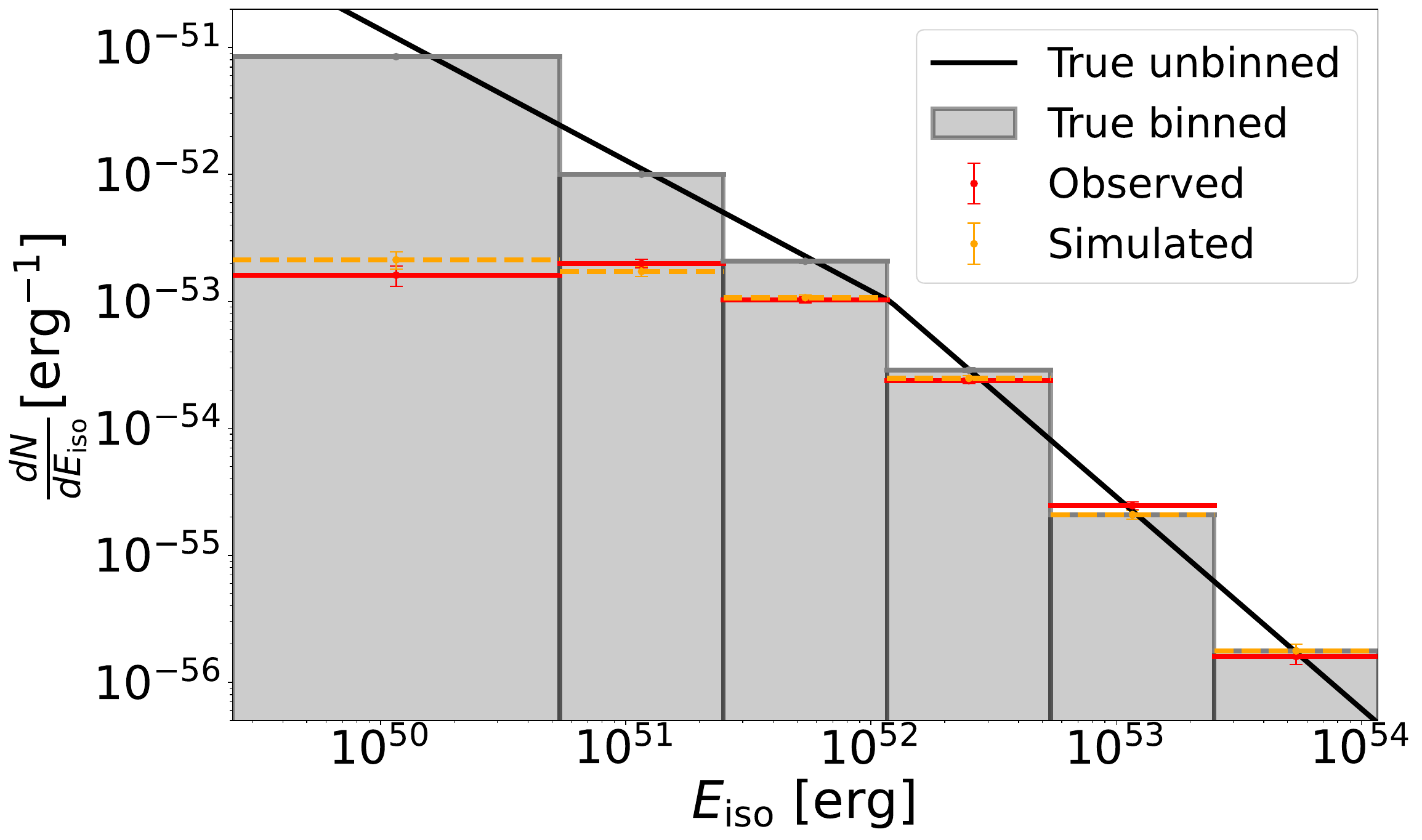}
 \caption{The black solid line represents the best-fit BPL intrinsic differential energy distribution $dN/dE_{\rm iso}$ obtained for the full sample. The gray shaded histogram is a binned version of the same distribution. The orange distribution was simulated from the intrinsic one after accounting for the selection effects, to be compared with the red distribution, which is the observed one.}
\label{fig:BPL_best_fit_energy}
\end{figure}

Confidence intervals were found by considering regions of the parameter space that satisfy $p_{\alpha} > 1 - \alpha $, where $\alpha$ is the confidence level, and  $p_{\alpha}$ is the p-value of the statistical test (either KS, AD, or $\chi^2$ test). We took $\alpha = 0.95$. Fit goodnesses (estimated with the p-values) and confidences intervals for the different models, either for the full sample or for the GS sample, are reported in Table \ref{table:pvalue_best_fit_E_iso}. 
Analogous conclusions were obtained for the GS, that is a PL distribution cannot account for the data, whereas a BPL with similar shape as the one obtained for the whole sample does. Hence, the need for a break in the intrinsic distribution does not depend on the quality of the modeling of the GRB light curves, at least as long as $dN/dE_{\rm iso}$ is concerned.
Figure~\ref{fig:BPL_best_fit_energy} shows the best fit model for $dN/dE_{\rm iso}$ obtained for the full sample.

\begin{table*}
\centering
\begin{tabular}{||l|c|c|c|c|c|c||c|c||} 
 \hline
 Model  & Data set & $s_1$ & $s_2$ & $E_b~(10^{51} \rm{erg})$ & KS test p-value& AD test p-value & $(\chi^{2}$/dof~;~p-value)\\ [0.5ex] 
 \hline\hline
 
 PL  & all & $\sim 1.4$ & - & - & $10^{-5}$ & $<10^{-3}$ & $(60/5~;~10^{-11}$)\\[0.1cm]
 PL  & all (second model) & $\sim 1.4$ & - & - & $7 \times 10^{-5}$ & $<10^{-3}$ & $(75/5~;~10^{-14}$)\\[0.1cm]
 BPL  & all & [0] & $\sim 1.5$ & $\sim 2$ & $0.096$ & $0.015$ & $(26.8/5~;~6 \times 10^{-5})$\\[0.1cm]
 BPL  & all & $\sim 1$ & [$1.5$] & $\sim 6$ & $0.17$ & $0.08$ & $(11.3/5~;~0.046)$\\[0.1cm]
 BPL  & all & $0.96^{+0.23}_{-0.15}$ &  $1.67^{+0.23}_{-0.16}$ & $12_{-11.4}^{+29}$ & $0.11$ & $0.14$ & $(5.25/5~;~0.39)$\\[0.1cm]
 BPL  & all (second model) & $1.02^{+0.10}_{-0.10}$ &  $1.79^{+0.08}_{-0.10}$ & $21_{-6}^{+21}$ & $0.23$ & $0.22$ & $(13/5~;~0.025)$\\[0.1cm]
 PL  & GS & $\sim 1.4$ & - & - &  $0.02$ & $0.002$ & $(27/5 ; 6 \times 10^{-5})$ \\[0.1cm]
 BPL  & GS & [0]& $\sim 1.4$ & $\sim 1$ &  $0.07$ & $0.025$ & $(14.8/5 ; 0.01)$ \\[0.1cm]
 BPL  & GS & $1.03^{+0.21}_{-0.27}$ & [$1.5$] & $11_{-10.6}^{+27}$ &  $0.23$ & $0.21$ & $(9.3/5 ; 0.097)$ \\[0.1cm]
 BPL  & GS & $1.07^{+0.16}_{-0.03}$ & $1.68^{+0.38}_{-0.11}$ & $24_{-7}^{+59}$ &  $0.2$ & $0.25$ & $(5.6/5 ; 0.34)$ \\[0.1cm]
  PL & $z<1.76$ & $\sim 1.4$ & - & - & $10^{-3}$ & $<10^{-3}$ & ($47.5/5$; $4 \times 10^{-9}$) \\[0.1cm]
   BPL  & $z<1.76$ & [0] & $1.48^{+0.14}_{-0.02}$ & $0.93_{-0.16}^{+0.86}$ &  $0.3$ &  $0.25$ & $(11/5 ; 0.05)$ \\[0.1cm]
   BPL  & $z<1.76$ & $0.82^{+0.10}_{-0.19}$ & [$1.5$] & $2.85^{+1.28}_{-1.79}$ &  $0.28$ & $0.25$ & $(9.87/5 ; 0.08)$ \\[0.1cm]
    BPL  & $z<1.76$ & $0.77^{+0.41}_{-0.39}$ & $1.65^{+0.35}_{-0.15}$ & $4.9^{+24}_{-3.2}$ &  $0.6$ & $0.53$ & $(3.6/5 ; 0.6)$ \\[0.1cm]
  PL & $N_p \leq 6$ & $\sim 1.4$ & - & - & $0.10$ & $0.02$ & $(17/5; 4 \times 10^{-3})$ \\[0.1cm]
   PL & $N_p > 6$ &$\sim 1.5$ & - & - & $5 \times 10^{-3}$ & $<10^{-3}$ & $(50/5; 10^{-9})$ \\[0.1cm]
   BPL & $N_p > 6$ &[0] & $\sim 1$ & $\sim 1.5$ & $0.38$ & $0.17$ & $(14/5 ; 0.01)$ \\[0.1cm]
   BPL & $N_p > 6$ &$\sim 0.7$ & $[1.5]$ & $\sim 1.5$ & $0.35$ & $0.17$ & $(11.5/5 ; 0.04)$ \\[0.1cm]
    BPL & $N_p > 6$ & $0.73_{-0.15}^{+0.23}$ & $ 1.54_{-0.03}^{+0.08}$ & $2.7_{-0.5}^{+1.7}$ & $0.5$ & $>0.25$ & $(9.2/5 ; 0.1)$ \\[0.1cm]
 \hline
\end{tabular}
\caption{Results of the modeling of the $dN/dE_{\rm iso}$ distribution, once selection effects are accounted for. Values of frozen parameters are reported in square brackets.}
\label{table:pvalue_best_fit_E_iso}
\end{table*}

\begin{figure}
 \includegraphics[width=\columnwidth]{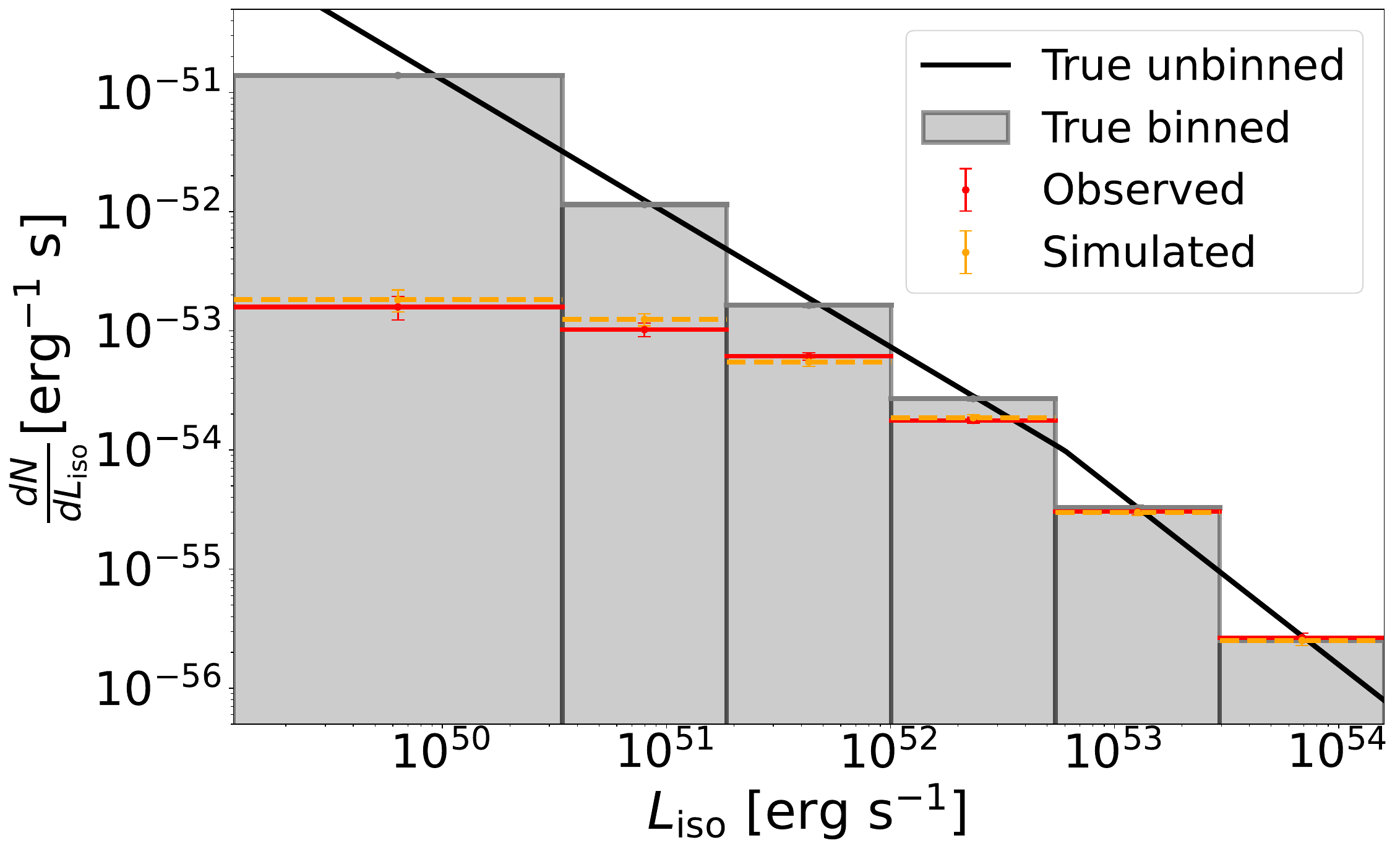}
 \caption{Same as Figure~\ref{fig:BPL_best_fit_energy}, except that here it is the differential luminosity distribution $dN/dL_{\rm iso}$.}
\label{fig:BPL_best_fit_luminosity}
\end{figure}

We obtained similar results for $dN/dL_{\rm iso}$ for the full sample, i.e a simple PL distribution is unable to describe the observed distribution (the best fit is obtained around $\kappa \sim 1.3$ with $\chi_{5}^2/5 \sim 20/5$), while a BPL works better (the best fit is obtained around $\kappa_1 \sim 1.1$, $\kappa_2 \sim 1.5$, $L_b \sim 6 \times 10^{52}$~erg~s$^{-1}$ with $\chi_{5}^2/5 \sim 3.4/5$). For the GS sample, the observed distribution is compatible with a PL distribution, the BPL distribution leading only to a marginal improvement. Results regarding $L_{\rm iso}$ distribution, either for the full sample or for the GS sample, can be found in Table \ref{table:pvalues_best_fit_Liso}. Figure \ref{fig:BPL_best_fit_luminosity} shows the BPL that best fits $dN/dL_{\rm iso}$ of the full sample.

\begin{figure}
 \includegraphics[width=\columnwidth]{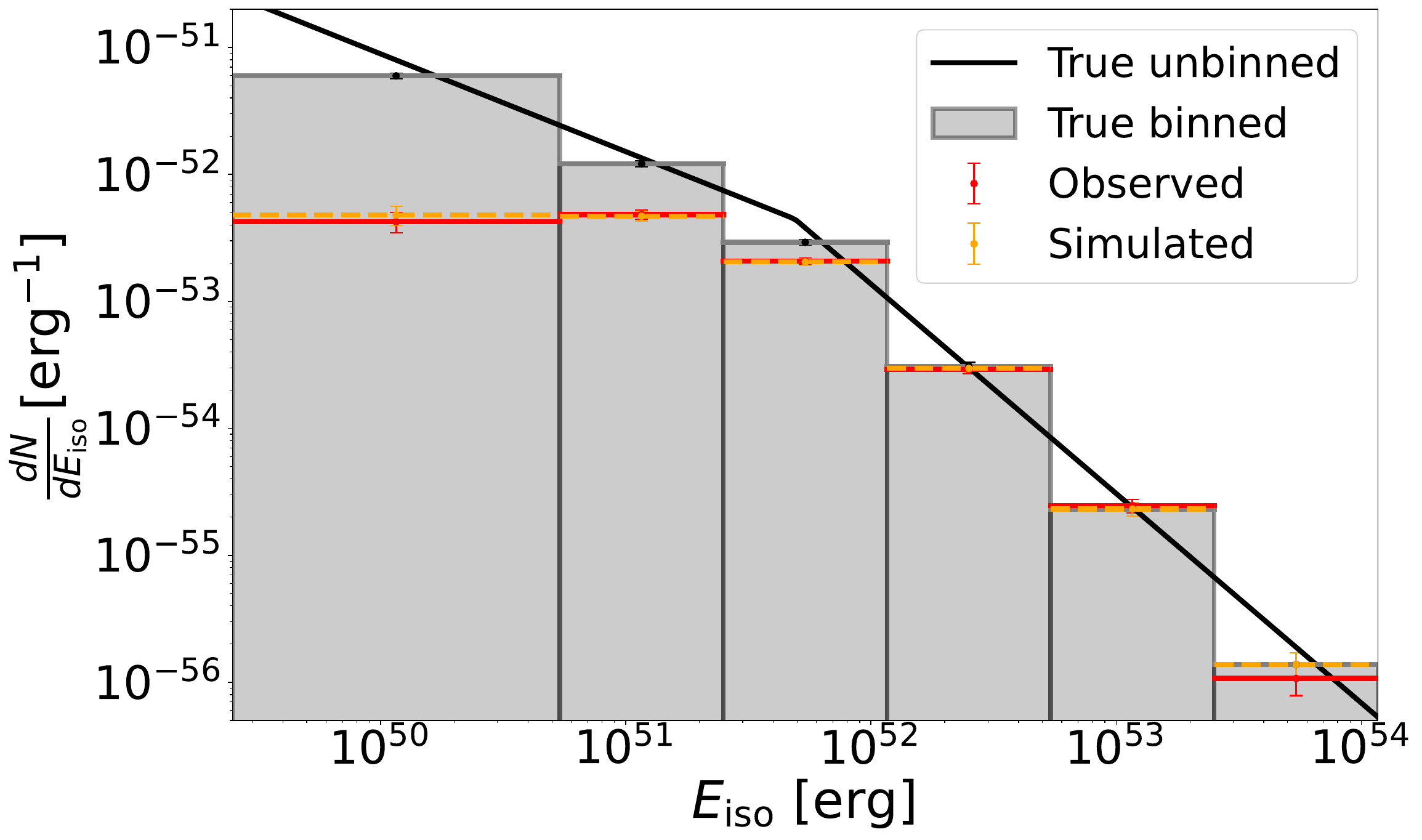}
 \caption{Same as Figure~\ref{fig:BPL_best_fit_energy}, except that here we are considering only pulses with redshift $z<1.76$.}
\label{fig:BPL_best_fit_low_z}
\end{figure}
\begin{table*}
\centering
\begin{tabular}{||c|c|c|c|c|c|c|c||} 
 \hline
 Model & Data set & $\kappa_1$ & $\kappa_2$ & $L_b~\rm{(10^{52}~erg~s^{-1})}$ &KS test p-value& AD test p-value & $(\chi^{2}$/dof~;~p-value)\\ [0.5ex] 
 \hline\hline
 PL & all &  $\sim 1.3$ & - & - & $0.04$ & $0.003$ & $(20/5~;~10^{-3})$  \\[0.1cm]
 PL & all (second model) &  $\sim 1.4$ & - & - & $0.01$ & $0.001$ & $(37/5~;5 \times 10^{-7})$  \\[0.1cm]
 BPL & all & $1.10_{-0.34}^{+0.07}$ & $1.47^{+0.60}_{-0.18}$ & $6_{-5.8}^{+14}$  &$0.33$ & $0.17$ & $(3.4/5~;~0.63)$ \\[0.1cm]
 BPL & all (second model) & $1.03^{+0.05}_{-0.06}$ & $1.51^{+0.05}_{-0.07}$  & $1.1^{+0.9}_{-0.6}$& $0.12$&$0.03$  &$(17/7~;~0.02)$ \\[0.1cm]
  PL & GS &  $1.32_{-0.05}^{+0.06}$ & - & - & $0.23$ & $0.09$ & $(9/5~;~0.1)$ \\[0.1cm]
    BPL & GS &  $1.17^{+0.13}_{-0.10}$ &  $1.47_{-0.06}^{+0.33}$ & $3_{-1}^{+25}$ & $0.74$ & $>0.25$ & $(2.8/5~;~0.73)$\\[1ex]
    PL & $z < 1.76$ &  $\sim 1.3$ & - & - & $10^{-3}$ & $10^{-3}$ & $(46/5~;~10^{-8})$\\[0.1cm]
    BPL & $z <1.76$ & $0.93^{+0.20}_{-0.21}$ & $1.73^{+0.64}_{-0.26}$ & $2.3^{+5.1}_{-1.2}$ & $0.7$ & $>0.25$ & $(2.5/5 ; 0.77)$\\[0.1cm]
     PL & $N_p \leq 6$ & $1.65^{+0.16}_{-0.16}$ & - & - & $0.5$ & $>0.25$ & $(6.9/5; 0.23)$ \\[0.1cm]
    PL & $N_p > 6$ & $\sim 1.2$ & - & - & $0.025$ & $< 10^{-3}$ & $(32/5 ; 6 \times 10^{-6})$ \\[0.1cm]
    BPL & $N_p > 6$ & $1.03^{+0.05}_{-0.11}$ & $1.73^{+0.64}_{-0.26}$ & $15^{15}_{-14.3}$ & $0.5$ & $>0.25$ & $(5.3/5 ; 0.38)$ \\[0.1cm]    
 \hline
\end{tabular}
\caption{Results of the modeling of the $dN/dL_{\rm iso}$ distribution, once selection effects are accounted for.}
\label{table:pvalues_best_fit_Liso}
\end{table*}

The need for a break in the $E_{\rm{iso}}$ distribution could be either real or due to our possible inability to model very accurately the selection effects. To verify the latter possibility, we carried out the same analysis after excluding the three redshift bins with the highest redshift, which are most severely affected by the selection effects. As a result, we limited to pulses from GRBs with $z_{\rm{max}} < 1.76$, which make up $\sim$$60$\% of the entire sample. Our conclusions remained essentially unchanged: the PL model is rejected (p-values from KS, AD, and $\chi^2$ tests were $8\times 10^{-3}$, $< 10^{-3}$, $10^{-11}$ respectively), whereas the BPL model was acceptable, with best-fit values fully compatible with those obtained for the full sample. Figure \ref{fig:BPL_best_fit_low_z} shows the BPL that best fits $dN/dE_{\rm iso}$ of the sample of pulses with $z<1.76$.

It is worth noting that we did not add pulses missed by {\sc mepsa} in the simulated light curves, as we did in Section~\ref{sec:fit} in the case of the real ones. 
Consequently, the estimated detection efficiency in the low tail of the distribution is somewhat lower than the effective one acting on real data. This further supports the evidence for a break ``a fortiori''.

%-----------------------------------------------
\subsubsection{Comparison between pulse-rich and pulse-poor GRBs}
\label{sec:pulse_rich_poor}
\begin{figure*}
 \includegraphics[width=1.\textwidth]{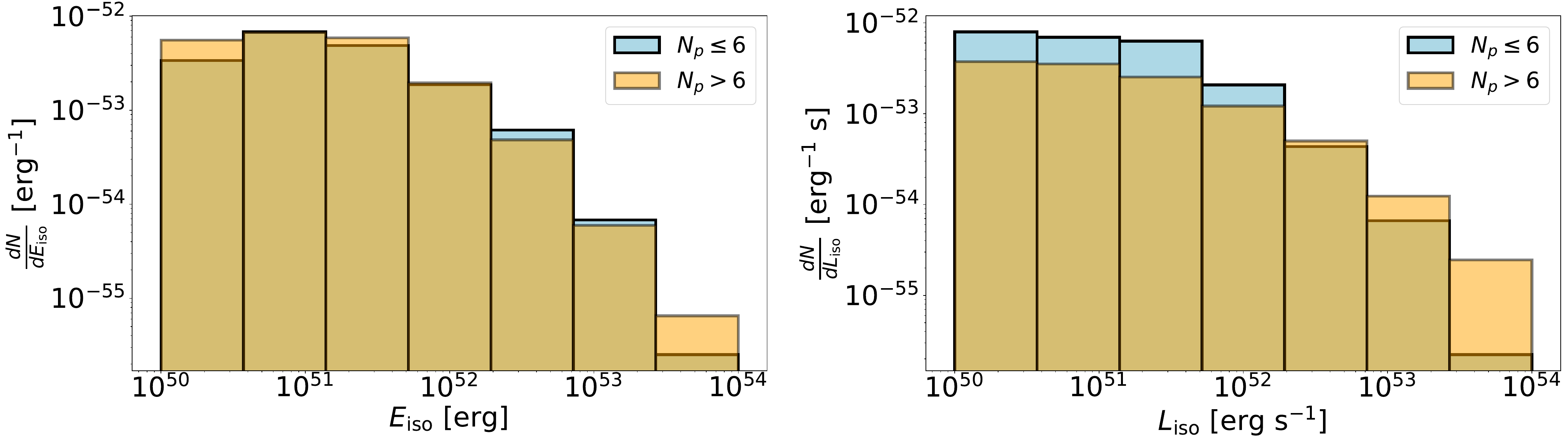}
 \caption{{\em Left} ({\em right}) $dN/dE_{\rm{iso}}$ ($dN/dL_{\rm{iso}}$) distributions for pulses belonging to pulse-poor GRBs (cyan) and to pulse-rich GRBs (orange).}
\label{fig:pulse_poor_vs_pulse_rich}
\end{figure*}

%-----------------------------------------------
We studied separately pulse-rich ($N_p>6$) and pulse-poor ($N_p \leq 6$) GRBs. The separation value was chosen so that the number of pulses belonging to pulse-poor GRBs is large enough to enable a statistical analysis. To this aim, we chose 6 pulses as the boundary, which corresponds to $\sim 20\%$ of the whole sample, that is 220 pulses, belonging to the pulse-poor GRBs. 
In the case of $dN/dE_{\rm{iso}}$, a two-population KS (AD) test between the two subsamples yielded a p-value of $0.3$ ($>0.25$), thus providing no evidence for a different parent population. On the contrary, for $dN/dL_{\rm{iso}}$ the same tests yielded a p-value of $10^{-15}$ ($<0.001$), thus rejecting a common parent distribution for the two groups. As shown in Figure~\ref{fig:pulse_poor_vs_pulse_rich}, least luminous pulses in pulse-poor GRBs are relatively more abundant than in pulse-rich GRBs, although being comparably energetic, suggesting that they are on average less luminous and longer. As was noted in Section~\ref{sec:validation_distrib}, the SS is predominantly populated by pulse-rich GRBs; thus, the difference in the luminosity distribution that was pointed out in the comparison between GS and SS is due to the pulse richness here considered.
%

%-----------------------------------------------
\subsubsection{Populations of pulses within individual GRBs}
\label{sec:lumpulses_within_GRBs}
%-----------------------------------------------
Thus far we studied the population of pulses as a whole, regardless of which GRB a given pulse may belong to.
We explored more in detail this aspect, by testing whether pulses distribute among different GRBs completely randomly, starting from the overall sample. To this aim, we considered the distribution of peak luminosity of GRBs (that is, the luminosity of the most prominent pulse within each GRB) and focused on individual redshift bins, to limit the impact of selection effects. 

\begin{figure}
 \includegraphics[width=\columnwidth]{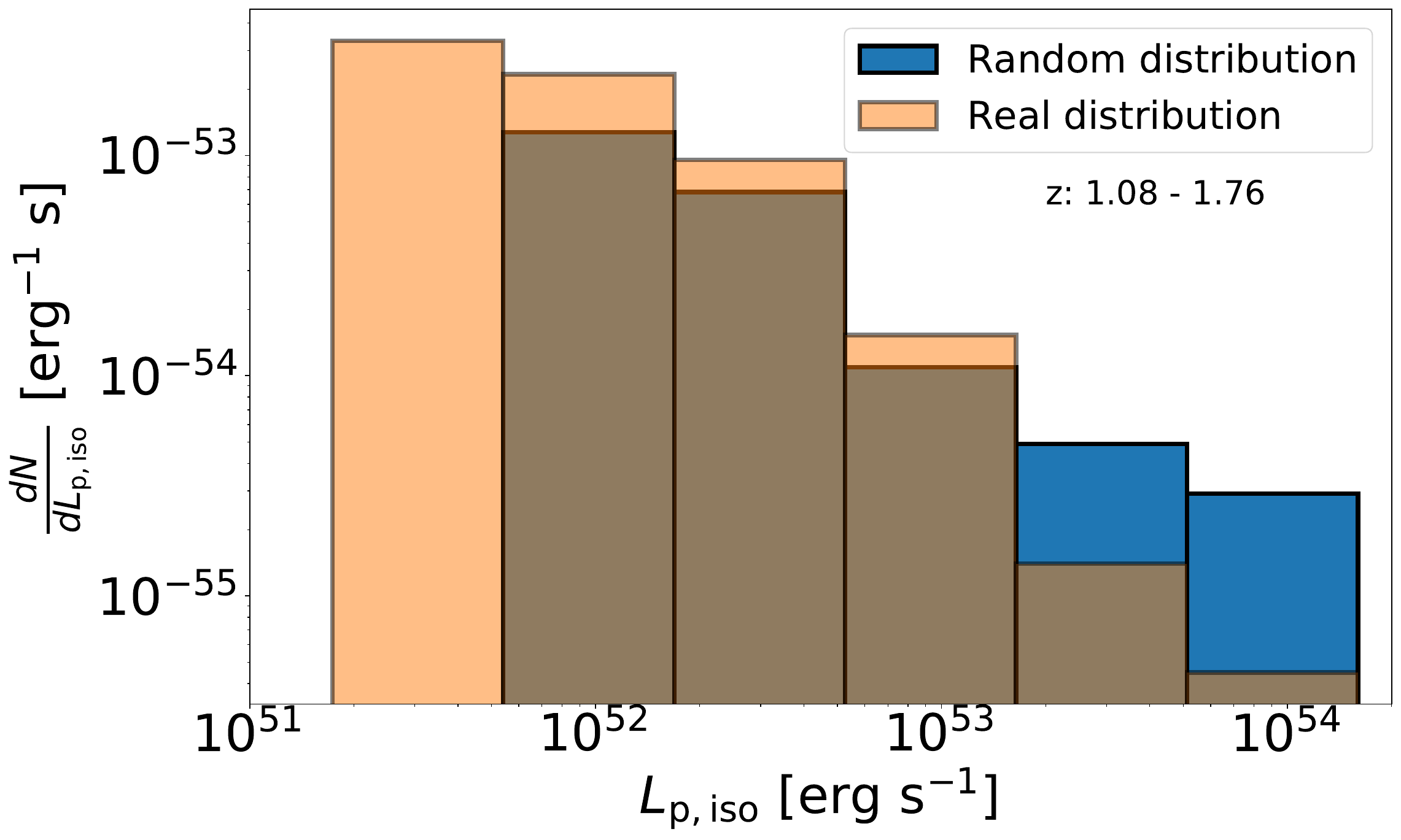}
 \caption{GRB peak luminosity distributions in the redshift bin $1.08<z<1.76$: real (orange) vs. randomly simulated (blue). The latter was obtained assuming that all pulses distribute randomly among the different GRBs. The harder tail of the simulated distribution suggests that luminous pulses belong to relatively few GRBs. See Section~\ref{sec:lumpulses_within_GRBs}.}
\label{fig:peak_lum_vs_mixed}
\end{figure}

For the $1.08<z<1.76$ redshift bin, which includes most GRBs ($N_{\rm grb}=41$, collecting $N_{p}=287$ pulses), we computed the GRB peak luminosity distribution. We then compared it with a distribution obtained by mixing together all the pulses belonging to the GRBs in the same redshift bin in the following way: for each GRB ($i=1,\ldots,N_{\rm grb}$) we drew a random set of $N_{p,i}$ pulses, where $N_{p,i}$ is the real number of pulses of GRB $i$ and took the peak luminosity of simulated GRB $i$. Repeating this for all of the $N_{\rm grb}$ GRBs, we obtained a fake peak luminosity distribution, which appears to be significantly different from the real one (AD test p-value $<10^{-3}$ that both sets share a common parent population): in particular, the fake distribution has a more populated hard tail (see Figure~\ref{fig:peak_lum_vs_mixed}). This indicates that pulses belonging to a given GRB are not completely independent of one another, or, at least, that they are not the result of randomly assembling from the whole population of observed pulses. High luminosity pulses, in particular, tend to cluster within a few GRBs, rather than being randomly distributed among all the GRBs.

To further test this possibility, we performed a multinomial test. Firstly, we counted $N_{\rm grb,lum}=4$ GRBs that contain the top 10\% most luminous pulses ($n=28$ out of 287 pulses). We then compared this number with the one that would be obtained by randomly distributing the pulses among all the GRBs, keeping the same distribution of number of pulses per each GRB. This problem is similar to simulate the outcomes of $n$ throws of a dice with $m$ faces and can be addressed by sampling from the multinomial distribution, described by its probability mass function:\footnote{We made use of {\tt scipy.stats.multinomial.rvs} to draw samples from the multinomial distribution.}
\begin{equation}
     {\rm PMF}({\bf n}\,|\, {\bf p}, m, n)\ =\  n!\prod_{i=1}^{m}\frac{p_{i}^{n_{i}}}{n_i!}\;,
\label{eq:PMF}
\end{equation}
where ${\bf n}=(n_1, n_2, \ldots, n_m)$ is the array of number of luminous pulses in each GRB ($\sum_{i=1}^m n_i = n = 28$), $m = N_{\rm grb} = 41$, and ${\bf p} = (p_1, p_2, \ldots, p_m)$, with $p_i= N_{p,i}/N_p$ being the probability for a pulse to belong to GRB $i$, calculated from all $N_p=287$ pulses in the chosen redshift bin. We performed $10^{7}$ simulations and in all cases the number of GRBs that contained the luminous pulses was higher (four times in average) than the real one, $N_{\rm grb,lum}=4$. The p-value that luminous pulses randomly distribute among GRBs within this redshift bin is $<5\times10^{-5}$.
 
We replicated the same analysis on the next redshift bin, $1.76<z<2.90$,  which contains $N_{\rm grb}=25$ and $N_{p}=253$ pulses. The top 10\% luminous pulses ($n=25$) belong to 5 GRBs. The same test yielded a p-value of $0.9\times 10^{-4}$ that luminous pulses are distributed over $\le 5$ GRBs. We can therefore conclude that the most luminous pulses are more likely to belong to fewer GRBs than what is expected from a pure random distribution.

%-----------------------------------------------
\subsection{Waiting time and duration distributions}
\label{sec:waiting_times}
%-----------------------------------------------
We obtained the distribution of the rest-frame WT calculated as
$\Delta t_i = (t_{0,i+1} - t_{0,i})/(1+z)$ where $t_{0,i}$ is the pulse peak time of the $i^{\rm{th}}$ pulse. We also calculated the pulse duration distribution, where duration was defined as the full width at half maximum (FWHM) of the modeled pulse, $T_i = \tau_{ri}(1+\xi_i)\log(2)^{1/\nu_i}$. Durations were not corrected for cosmological dilation because of the combination of different effects that mostly cancel each other (see \citealt{Camisasca23} and references therein for a detailed explanation).

\begin{figure}
 \includegraphics[width=\columnwidth]{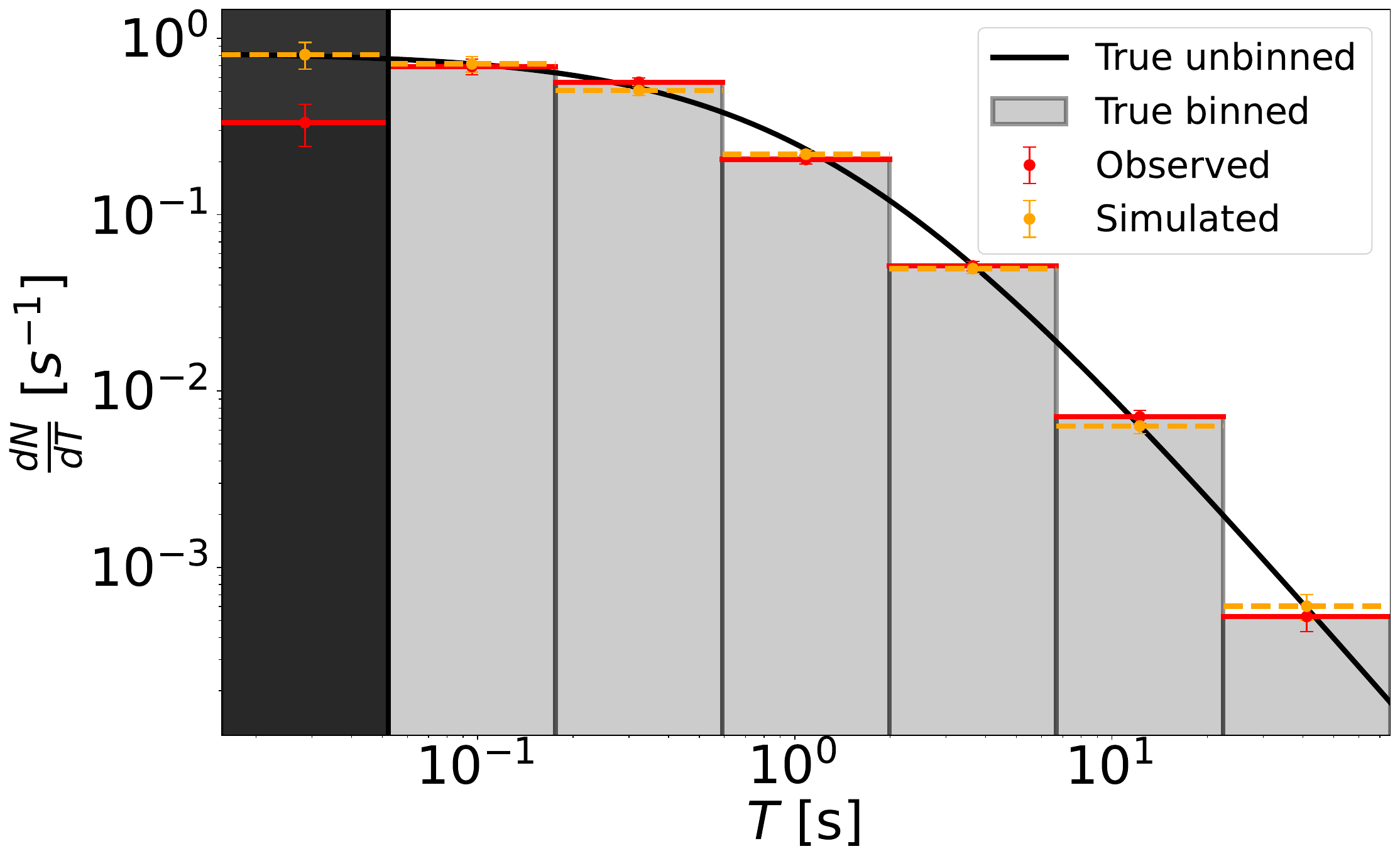}
 \caption{Same as Figure~\ref{fig:BPL_best_fit_energy}, except that here it is the differential duration distribution $dN/dT$. The black area represents the first bin which is not considered in the computation of the posterior distribution.}
\label{fig:time_dur}
\end{figure}

\begin{figure}
 \includegraphics[width=\columnwidth]{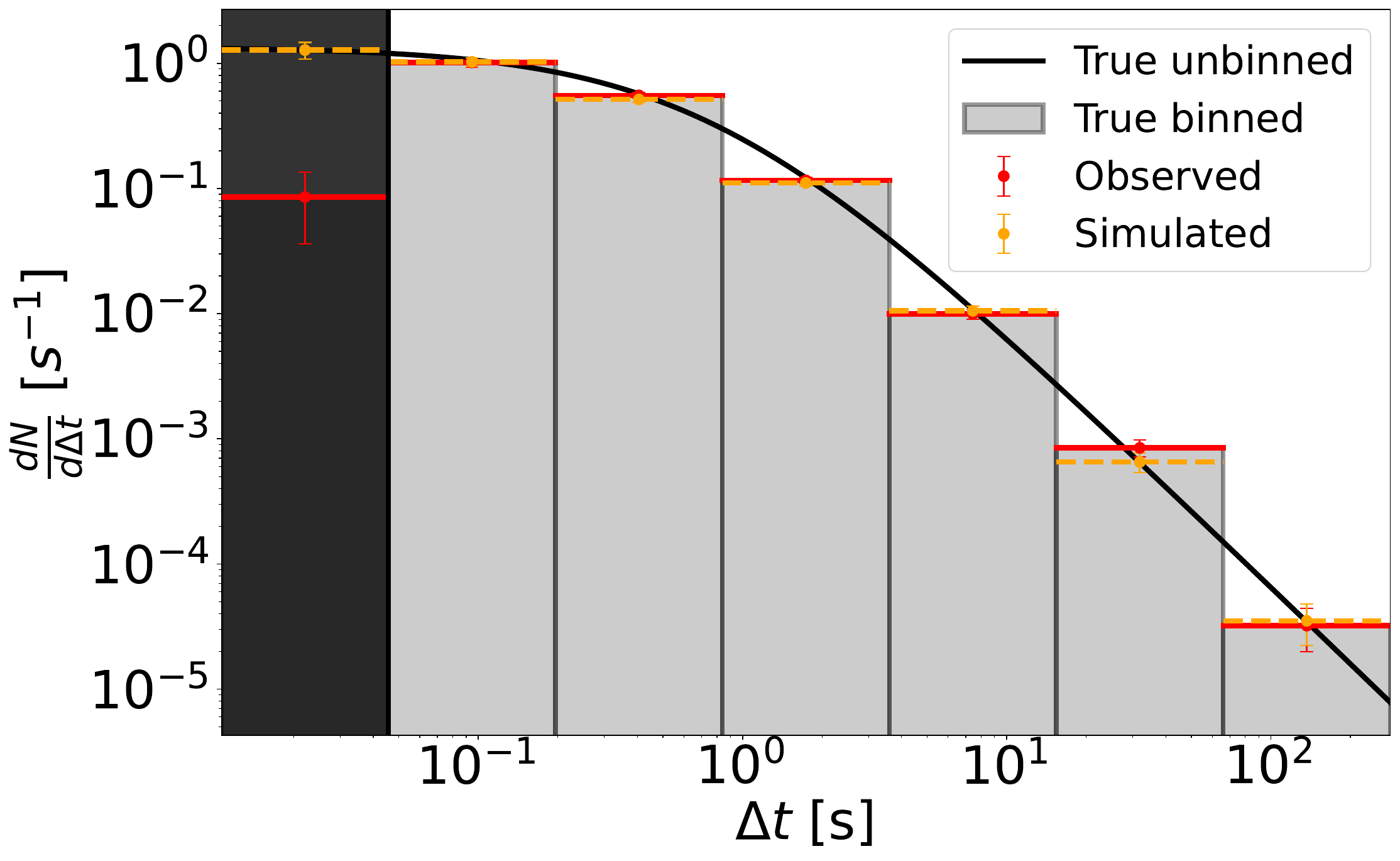}
 \caption{Same as Figure~\ref{fig:BPL_best_fit_energy}, except that here it is the differential duration distribution $dN/d\Delta t$. The black area represents the first bin which is not considered in the computation of the posterior distribution.}
\label{fig:wts}
\end{figure}

To ease the comparison with previous results, we modeled the distributions as in \citet{Guidorzi15b}, using the following model:
\begin{equation}
    \frac{dN}{d \Delta t} = (2-\gamma)\delta^{2-\gamma}(\delta+\Delta t)^{-(3-\gamma)}
\end{equation}

consisting of a PL hard-tail with a PL index $3-\gamma$, plus a characteristic break time $\delta$ below which the distribution flattens 
($\gamma$ and $\delta$ corresponds to $\alpha$ and $\beta$ in Equation 8 of \citealt{Guidorzi15b}). The first bin of the two distributions was not considered ($\Delta t \geq 0.05~\rm{s}$) since {\sc mepsa} efficiency drops at low WTs/durations. We sampled the posterior distribution by running a MCMC simulation with $10^{4}$ steps and 32 walkers. We discarded the $10^{3}$ first steps of the posterior distribution and we used only every 15 steps from the chain. For the duration and the WTs distribution, we obtained a PL index $3-\gamma_T = 2.08^{+0.19}_{-0.16}$ and $3-\gamma_{WT} = 2.04^{+0.14}_{-0.12}$, respectively.
Best fit parameters are reported in Table~\ref{table:wts} while the distributions along with the best fit models are shown in Figures~\ref{fig:time_dur} and \ref{fig:wts}.

\begin{table}
\centering
\begin{tabular}{||c|c|c||} 
\hline
   & $3-\gamma$ & $\delta~(s)$ \\
\hline\hline
  WT & $2.04^{+0.14}_{-0.12}$ &$0.78^{+0.18}_{-0.15}$\\[0.1cm]
  Pulse FWHM & $2 .08^{+0.19}_{-0.16}$& $1.30^{+0.37}_{-0.28}$\\[0.1cm]
\hline
\end{tabular}
\caption{Results of the modeling of the differential distributions $dN/d\Delta t$ and $dN/dT$. Uncertainties are given with a 90 $\%$ confidence level.}
\label{table:wts}
\end{table}
     
%---------------------------------------------
\subsection{Effects of the pulse model}
\label{sec:new_model}

\begin{figure}
 \includegraphics[width=\columnwidth]{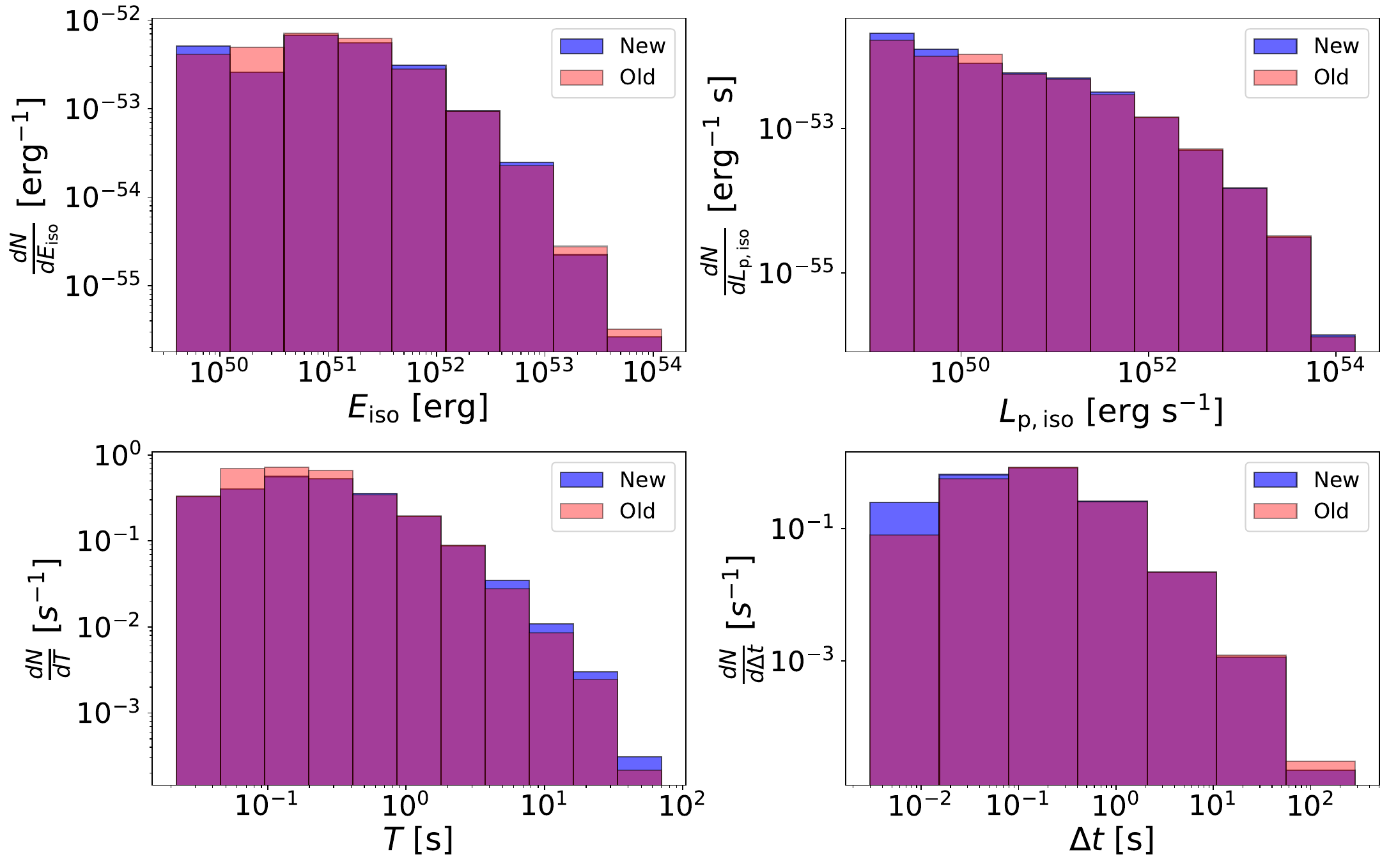}
 \caption{$E_{\rm{iso}}$ (upper left), $L_{\rm{iso}}$ (upper right), duration (lower left), and WT (lower right) distributions computed either with the first pulse model (Section~\ref{sec:fit}; in red) or the second (\citealt{Norris05}; in blue).}
 \label{fig:old_vs_new_distribs}
\end{figure}

We repeated the analysis by adopting the alternative pulse model of \citet{Norris05}, to explore to which extent our results are sensitive to the choice of the model described in Section~\ref{sec:fit}. Overall, the fit quality is worse than the one obtained with the former pulse model, since the number of GRB LCs that do not fulfill the condition $|\sigma_\epsilon-1| > 0.1$ (see Section~\ref{sec:results}) is 57 vs. former 15. This was to be expected because of the smaller number of model parameters.
We then computed the four different distributions considered in this paper and compared them with the ones obtained with the former pulse model (Figure~\ref{fig:old_vs_new_distribs}). 

We also performed two-population KS and AD tests. All but one test yielded p-values above 5\%, except for $0.3$\% obtained for the time duration distribution. This is due to the fact that the FWHM computed with the second model is on average longer (by a factor $\lesssim 2$) than the one computed with the first model.
The distributions were modeled following the same procedure adopted for the first pulse model. Overall, the power--law indices do not differ significantly from the previous corresponding ones: $\alpha^{(2)}_E = 1.79^{+0.08}_{-0.010}$, $\alpha^{(2)}_P = 1.51^{+0.05}_{-0.07}$,  $\alpha^{(2)}_T = 2.13^{+0.21}_{-0.08}$,  $\alpha^{(2)}_{WT} = 2.14^{+0.16}_{-0.14}$. Compared with previous values reported in Table~\ref{table:comparison_results}, they are all consistent. While the need for a break in energy/luminosity distributions is confirmed for both pulse models, in the luminosity case the value of the break point depends on the pulse model: for the second one, we found $L^{(2)}_{\rm{break}} \sim1\times 10^{52}~\rm{erg~s^{-1}}$, which deviates by $2.5\,\sigma$ from previous $L^{(1)}_{\rm{break}}\sim6\times10^{52}~\rm{erg~s^{-1}}$.
As a result, we proved that our results do not depend significantly on the pulse shape model, except for the value of the break luminosity, which we conservatively estimate in the range $L_b\sim10^{52-53}~\rm{erg~s^{-1}}$.

%---------------------------------------------

%%%%%%%%%%%%%%%%%%%%%%%%%%%%%%%%%%%%%%%%%%%%%%%%%%
\section{Discussion}
\label{sec:disc}
%%%%%%%%%%%%%%%%%%%%%%%%%%%%%%%%%%%%%%%%%%%%%%%%%%
For the distributions of energy, peak flux (or the corresponding intrinsic quantity which is luminosity), and duration, the corresponding PL indices predicted by SOC models are given by the following \citep{Aschwanden14b}:
\begin{equation}
    \alpha_E = 1+\frac{d-1}{D_{d}+\frac{2}{\beta}}
    \label{eq:alphaE}
\end{equation}
\begin{equation}
    \alpha_P = 1+\frac{d-1}{d}\;,
    \label{eq:alphaP}
\end{equation}
\begin{equation}
    \alpha_T = 1+\beta\frac{(d-1)}{2}
    \label{eq:alphaT}
\end{equation}
where $\beta$ is the spreading exponent ($\beta = 1/2$ for sub-diffusion, $\beta = 1$ for normal diffusion, $\beta = 2$ for linear expansion), $d$ is the Euclidean space dimension, and $D_{d}$ is the fractal dimension of SOC avalanches, which is usually approximated by $D_d \simeq \frac{1+d}{2}$. SOC theory also predicts $\alpha_{\rm{WT}} = \alpha_T$, where $\alpha_{\rm{WT}}$ is the PL index of the WT distribution (this holds for short WTs: for WTs longer than the longest pulse duration, the WT distribution should break exponentially; see \citealt{Aschwanden14b}).
In the case of 3D avalanches ($d=3$) and normal diffusion ($\beta=1$), it is $\alpha_E=1.5$, $\alpha_P=5/3$, and  $\alpha_T= \alpha_{\rm{WT}} = 2$.
%($\beta=1$),($d=3$)
Our estimates of the duration and of the WT duration distribution PL indices agree with SOC predictions: $\alpha_{\rm{T}} = 3-\gamma_T = 2.08^{+0.19}_{-0.16}$ and  $\alpha_{\rm{WT}} = 3-\gamma_{WT} = 2.04^{+0.14}_{-0.12}$. 

\begin{figure*}[h!]
 \includegraphics[width=\textwidth]{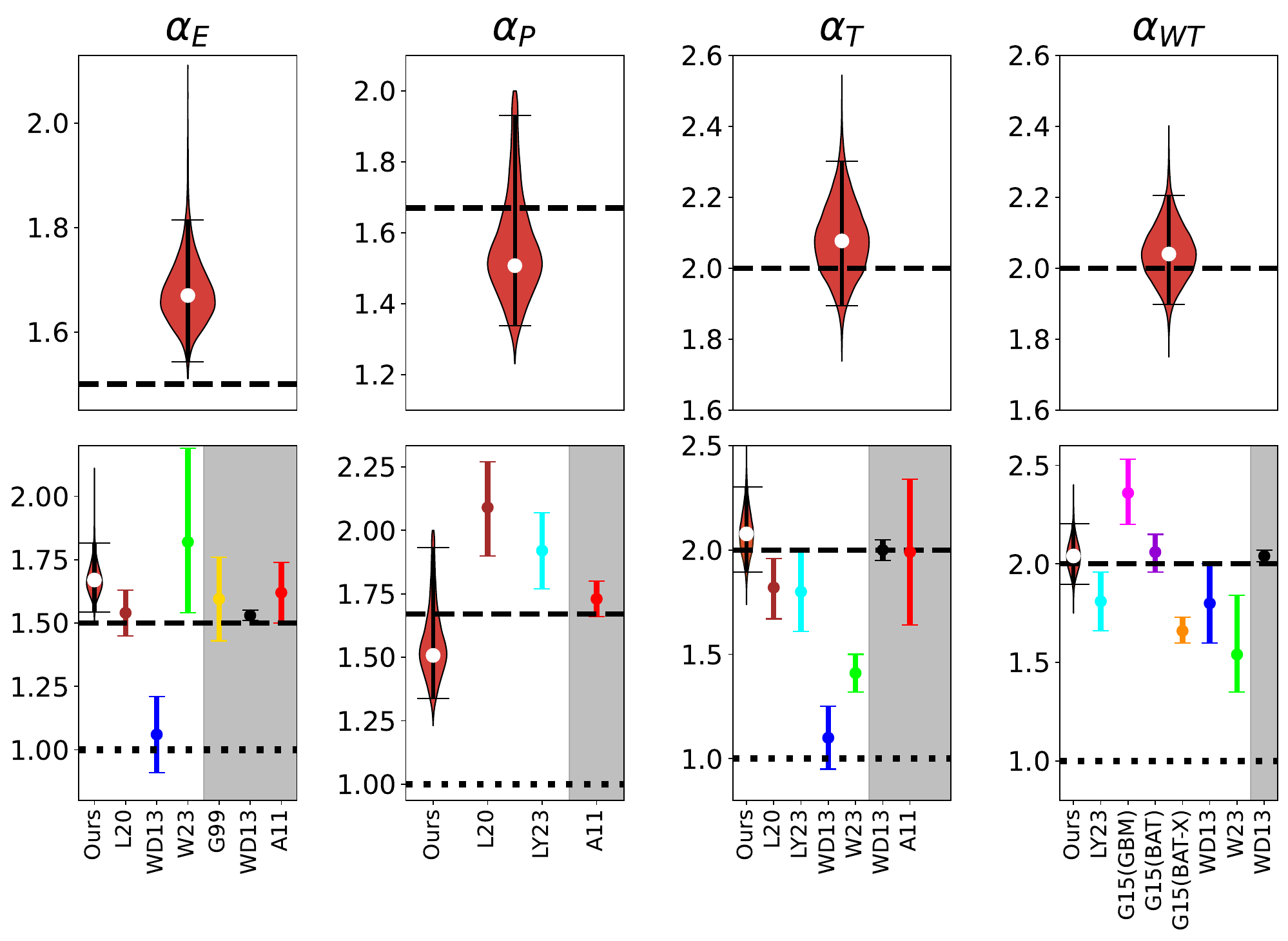}
 \caption{{\em Top panels, left to right}: violin plots of the posterior distributions for the PL indices of the energy, luminosity, duration, and waiting times distributions, respectively, obtained in this work on GRB prompt emission (for the energy and luminosity distributions, the post-break values were considered); the horizontal bars span 5 to 95\% quantiles. Dashed lines show the SOC predictions for $d=3$. {\em Bottom panels, left to right}: the same corresponding violin plots as in the top, including values reported from the literature. For GRB prompt emission: \citealt{Lyu20} (L20; brown), \citealt{Li23b} (LY23; cyan), and \citet{Guidorzi15b} (G15(GBM) and G15(BAT); in pink and purple). For GRB X-ray flares, values from \citealt{WangDai13} (WD13; blue), \citealt{Wei23} (W23; green). The orange point refers to a study where prompt emission pulses and X-ray flares were considered as a unique process, (\citealt{Guidorzi15b}; G15(BAT-X)). The shaded areas show the results reported in the literature from analogous investigations about different classes of astrophysical sources:
 for solar flares, values from \citealt{WangDai13} (WD13; black) and \citealt{Aschwanden11} (A11; red). For magnetars, values reported in \citealt{Gogus99,Gogus00} are displayed (G99; yellow). Dashed (dotted) lines show SOC predictions for $d = 3$ ($d=1$) and $\beta=1$.}
\label{fig:confrontoSOC}
\end{figure*}

The PL indices of the high-value tails of the energy and luminosity distributions, respectively $s_2 = 1.67^{+0.23}_{-0.16}$  and $\kappa_2=1.47^{+0.60}_{-0.18}$, are also roughly in agreement with SOC predictions (see Figure~\ref{fig:confrontoSOC}). 

However, our results show evidence for a break in both distributions, which is not predicted by SOC theory. Given the accuracy with which the detection efficiency was modeled, the evidence for a break seems to be hardly entirely ascribable to unaccounted selection effects.
In our study, we have selected all Type-II GRB candidates. Our study implicitly assumes that all Type-II GRB engines are behaving in the same way. However, one cannot reject the possibility that the observed populations is actually the mixture of different kinds of engines or at least different behaviors. Even for a given type of central engine, say an hyper accreting BH, two or more jet formation scenarios ($\nu\bar{\nu}$ annihilation or BZ effect) producing different energy/luminosity distributions, can be at play and contribute to the observed population of GRBs. In principle, the observed deviations from the SOC predictions could be also ascribed to the physics and geometry of the prompt emission, i.e., the properties of the jet and its interactions with the stellar envelope. Finally, a PL behavior does not unavoidably imply SOC, is just a necessary condition, not a sufficient one, given that hard-tailed distributions modeled as PL or BPL can be the result from a number of different processes.

Table~\ref{table:comparison_results} summarizes and compares our findings with analogous studies. Interestingly, the energy index obtained is consistent with what was found for the burst activity observed in Galactic magnetars: $\alpha_E = 1.43-1.76$ \citep{Gogus99,Gogus00}, and in solar flares: $\alpha_E = 1.62_{-0.12}^{+0.12}$ \citep{Aschwanden11}. We obtained substantially lower values for $\alpha_{P}$ than  \citet{Lyu20} and \citet{Li23b}. In these studies, $\alpha_P$ is determined by analyzing the distribution of pulse count rates given by the instrument, as it was also done for solar flares. While the latter case is justified, as all flares come from the same source at a fixed distance from the detector, for GRBs the range of distances is so large that one should either consider a group of GRBs with similar redshift or better use luminosities and model the selection effects that inevitably affect the observed sample.

We compared $dN/dL_{\rm iso}$ with the GRB (peak) luminosity function, as modeled for example in \citet{Ghirlanda22}. Even though the latter is in principle different from $dN/dL_{\rm iso}$, the two distributions are not completely independent of one another. \citet{Ghirlanda22} modeled the distribution with a BPL with low/high luminosity indices $a_1 \sim 1$ and $a_2 \sim 2.2$ and a break luminosity $L_b \sim 10^{52}(1+z)^{0.64}~\rm{erg~s^{-1}}$ slightly dependent on redshift. Our values are consistent with the low-luminosity index and with the break luminosity of the luminosity function. We find a flatter high-luminosity index that can be understood since our distribution contains all pulses of all GRBs, while the GRB peak luminosity function is determined by the most luminous pulse within each GRB. As we already pointed out, high-luminosity pulses tend to cluster in relatively few luminous GRBs: consequently, the distribution resulting from the selection of the most luminous pulse of each GRB turns into a depletion of luminous pulses and hence into a smaller fraction of high-luminosity events in the GRB distribution compared with the pulse distribution.

\begin{table*}
\centering
\addtolength{\tabcolsep}{-4pt}
\begin{tabular}{|c|cccc|cc|} 
\hline
 & $\alpha_{E}$ & $\alpha_{P}$ & $\alpha_{T}$ & $\alpha_{WT}$ & SOC $d=3$ &  SOC $d=1$\\[0.5ex] 
 \hline
GRB prompt emission & &  &  &  &  & \\[1ex]
Our study & $1.67^{+0.23}_{-0.16}$ & $1.47^{+0.60}_{-0.18}$ &   $2.08^{+0.19}_{-0.16}$ & $2.04^{+0.14}_{-0.12}$ & roughly & No \\
\citealt{Guidorzi15b} (GBM) & &  &  & $2.36^{+0.17}_{-0.16}$ & & \\
\citealt{Guidorzi15b} (BAT) &  &  &  & $2.06^{+0.10}_{-0.09}$ & & \\
\citealt{Lyu20} & $1.54^{+0.09}_{-0.09}$ & $2.09^{+0.18}_{-0.19}$ & $1.82^{+0.14}_{-0.15}$ &  & roughly & No\\
\citealt{Li23b} & & $1.92^{+0.15}_{-0.15}$  & $1.80^{+0.19}_{-0.19}$ &  & roughly & No \\ [0.1cm]
\hline
GRB X-ray flares &  &  &  &  & &\\ [0.5ex] 
\citealt{WangDai13} & $1.06^{+0.15}_{-0.15}$ &  & $1.10^{+0.15}_{-0.15}$ & $1.80^{+0.20}_{-0.20}$ &No & very roughly \\[0.1cm]
\citealt{Wei23} & $1.82^{+0.37}_{-0.28}$ & &$1.41_{-0.08}^{+0.09}$ & $1.54^{+0.30}_{-0.19}$ & No & No\\
\hline
GRB Prompt emission + X-ray flares & & &  &  & & \\[0.1cm]
\citealt{Guidorzi15b} (BAT-X) &  &  &  & $1.66^{+0.07}_{-0.06}$  & No&No \\[0.1cm]
\hline
GRB Precursors & & &  &  & & \\[0.1cm]
\citealt{Li23b} &   & $2.22^{+0.20}_{-0.20}$ & $1.82^{+0.19}_{-0.19}$ & $1.81^{+0.15}_{-0.15}$ & No&No \\[0.1cm]
\hline
SGRs &  &  & &  &  &  \\[0.1cm]
\citealt{Gogus99,Gogus00} & 1.43-1.76  &  &  &  & roughly&No \\[0.1cm]
\hline
Solar flares &  &  & &  &  &  \\[0.1cm]
\citealt{WangDai13} & $1.53^{+0.02}_{-0.02}$ & & $2.00^{+0.05}_{-0.05}$ & $2.04^{+0.03}_{-0.03}$ & Yes&No \\[0.1cm]
\citealt{Aschwanden11} & $1.62^{+0.12}_{-0.12}$ & $1.73^{+0.07}_{-0.07}$ & $1.99^{+0.35}_{-0.35}$ & & Yes&No \\[0.1cm]
\hline\hline
SOC predictions ($\beta=1$) & $3\frac{d+1}{d+5}$ &$2-\frac{1}{d}$& $\frac{d+1}{2}$  & $\frac{d+1}{2}$ & &\\[0.1cm]
$d =3$ & $\frac{3}{2}$ & $\frac{5}{3}$ & $2$ & $2$ & &\\[0.1cm]
$d =1$ & 1 & 1 & 1 & 1 & & \\[0.1cm]
\hline
\end{tabular}
\caption{Energy, luminosity, duration, and waiting times PL indices obtained in our study and in the literature in the case of GRB prompt emission, X-ray flares, precursors, solar flares, and SGRs. Values predicted by SOC theory are also indicated, in the case where $\beta=1$. Special cases of $d=1$ or $d=3$ are also reported. The last two columns indicate whether the obtained values are in agreement with the values predicted by SOC theory in the two special cases.}
\label{table:comparison_results}
\end{table*}

Finally, our results do not seem to crucially depend on redshift, given that they did not change in essence when we ignored GRBs with redshift  $z>1.76$. Specifically, the evidence for a break in both distributions holds true in both cases, suggesting that the break is not likely to be entirely due to selection effects, but it is an intrinsic feature.

%%%%%%%%%%%%%%%%%%%%%%%%%%%%%%%%%%%%%%%%%%%%%%%%%%
\section{Conclusions}
\label{sec:conc}
%%%%%%%%%%%%%%%%%%%%%%%%%%%%%%%%%%%%%%%%%%%%%%%%%%
We determined the energy, luminosity, duration, and waiting time distributions of individual pulses identified with a well-calibrated algorithm from 142 Type-II GRBs with known redshift detected by the GBM.  We then carried out a careful analysis of the selection effects through a suite of simulations, dividing our sample into 9 redshift bins, and modeled for each of them the detection efficiency as a function of energy (luminosity), $\eta_z(E_{\rm iso})$ ($\eta'_z(L_{\rm iso})$). For each redshift bin, we have then generated energy/luminosity samples of pulses accounting for the detection efficiency, under the assumption of some putative distribution models like simple PL or BPL. The plausibility of each assumed distribution was finally tested by comparing the resulting distribution with the observed one.

We found that a simple PL can reproduce neither $E_{\rm{iso}}$ nor $L_{\rm{iso}}$ distributions (especially in the case of $E_{\rm{iso}}$). Rather, for the first time we found evidence for a break in $E_{\rm{iso}}$ and $L_{\rm{iso}}$ distributions ($E_b \sim 10^{52}~\rm{erg}$ and $L_b~\sim~ 10^{52-53}~\rm{erg~s^{-1}}$), which appears to be hardly ascribable to unaccounted selection effects. A possible interpretation for this break is that the underlying assumption of a unique stochastic process ruling the GRB dynamics is not valid, but instead different kinds of dynamics, possibly connected with different progenitors or different regimes, contribute to the observed population of GRBs.

Interpreting our results in the SOC framework with $\beta=1$, our results are compatible with prompt emission being the result of 3D ($d=3$) avalanches produced by a non-linear dynamical system driven slowly to a SOC state.The SOC interpration of GRB prompt emission could be compatible with utterly different theoretical models. Indeed, this scenario could be consistent either with the picture of a NDAF surrounding a hyper-accreting stellar mass BH becoming thermally unstable and cooled by neutrino emission, with rotational energy being extracted by BZ effect and converted into a Poynting-flux dominated outflow, or with cascades of magnetic reconnection events produced within a magnetically dominated relativistic outflow at the GRB emission site, as foreseen in the ICMART model \citep{ICMART}.

\begin{acknowledgments}
 We acknowledge the anonymous reviewer for a constructive report which improved the manuscript.
R.M. and C.G. acknowledge the Dept. of Physics and Earth Science of the University of Ferrara for the financial support through the ``FIRD 2022'' grant. R.M. acknowledges the University of Ferrara for the financial support of his PhD scholarship. A.T. acknowledges financial support from ``ASI-INAF Accordo Attuativo HERMES Pathinder operazioni n. 2022-25-HH.0''.
\end{acknowledgments}

%%%%%%%%%%%%%%%%% APPENDICES %%%%%%%%%%%%%%%%%%%%%

\appendix

%-----------------------------------------
\section{Time-resolved spectral analysis}
\label{sec:time-resolved}
%-----------------------------------------

We computed the isotropic-equivalent released energy using Equation~\eqref{eq:eiso_time_avgd}, thus assuming a negligible impact of a possible spectral evolution.

GRB spectra are known to be temporally evolving. This is usually described in terms of two alternative behaviors: (i) a monotonic hard-to-soft evolution and (ii) pulse tracking (see~\citealt{Lu12b} and references therein). Therefore, the isotropic-equivalent energy should be computed using the time-resolved fluence $\Phi'_k$ instead of the time-average $\Phi_k$ used in Equation~\eqref{eq:eiso_time_avgd}, where $\Phi'_k$ is calculated from modeling the spectrum extracted within the time interval that includes only the $k$-th pulse (or, at least, centered on it, given that pulses occasionally overlap and therefore their spectra cannot be completely separated).

We conveniently defined the time-resolved conversion factor $f_k = \Phi'_k/C'_k$, where $C'_k$ are the counts integrated over the same time interval used to compute $\Phi'_k$. In this way, the degree of approximation introduced in Equation~\eqref{eq:eiso_time_avgd} by assuming time-average instead of time-resolved fluences can be studied through the ratio $f_k/f_{\rm tot}$.

To this aim, we considered a few cases of bright GRBs with numerous pulses, some of them being already known to show spectral evolution.

We modeled time-resolved spectra of these GRBs and we studied the impact of spectral evolution on the $E_{\rm{iso}}$ distribution. For this study, we considered 5 GRBs among our sample\footnote{They are 171010A, 180720B, 190114C, 170405A, and 090424.} having respectively 57, 34, 21, 20 and 13 pulses thus totaling 145, that is enough to build a statistically sound distribution. These GRBs are among the brightest of the GBM catalog, with three of them ranking in the 12 top fluence of the whole catalog.\footnote{171010A with $6.3\times10^{-4}$~erg~cm$^{-2}$; 190114C with $4.4\times 10^{-4}$~erg~cm$^{-2}$, and 180720B with $3.0\times10^{-4}$~erg~cm$^{-2}$.}

\begin{figure}
 \includegraphics[width=\columnwidth]{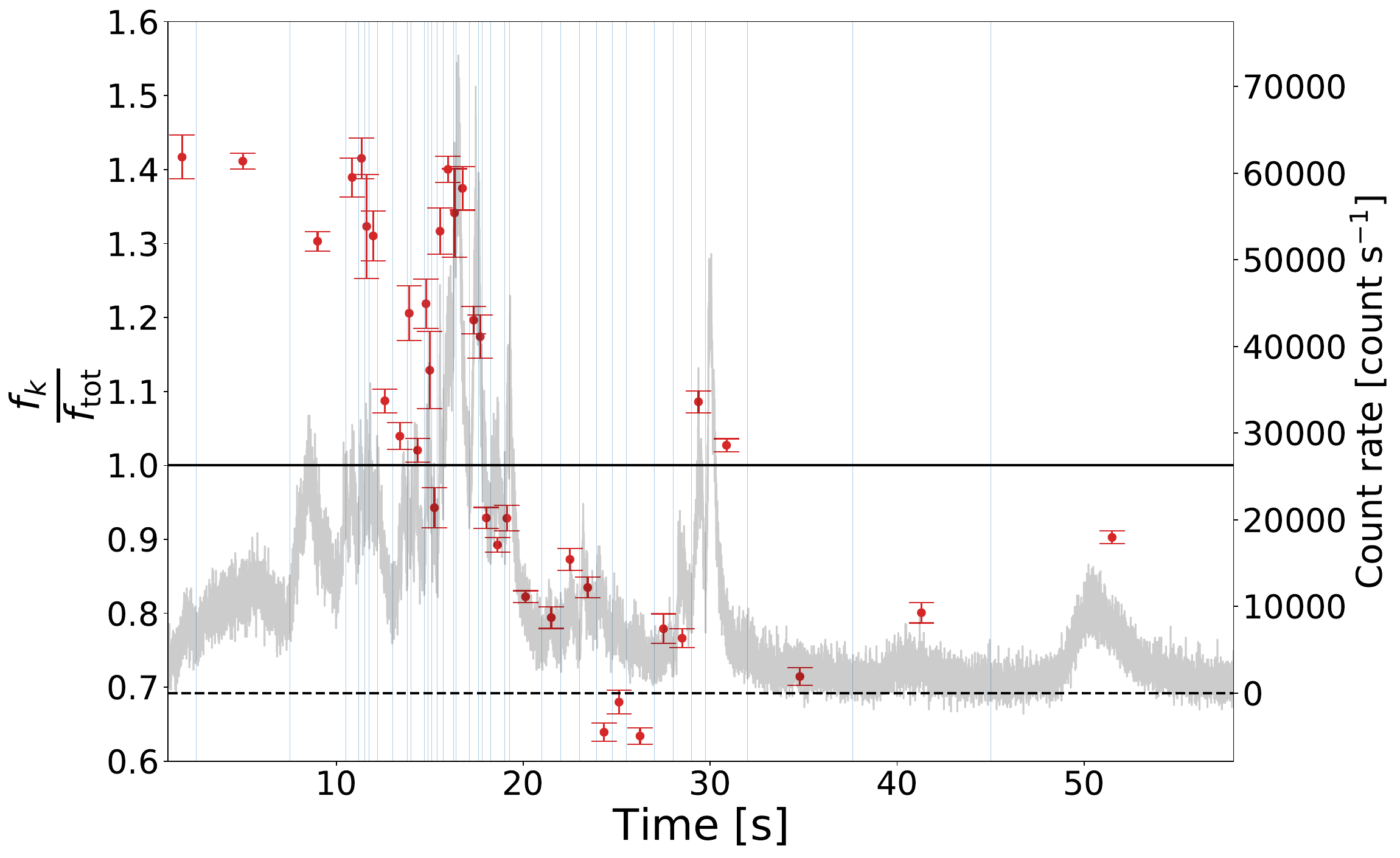}
 \caption{Evolution of the ratio of the time-resolved over time-average counts-to-fluence conversion factor, $f/f_{\rm tot}$ (shown by the red points), as a function of time for 180720B (the horizontal solid line shows the $f=f_{\rm tot}$ case). Blue vertical lines mark the boundaries of adjacent temporal intervals. The light curve is displayed in gray (right-hand $y$-axis).}
  \label{fig:180720B}
\end{figure}

Figure~\ref{fig:180720B} illustrates the case of 180720B: for the time-averaged spectrum, we took the fluence from the GRB catalog $\Phi_{\rm tot} = (2.9853 \pm 0.0008)\times10^{-4}$~erg~cm$^{-2}$, integrated from $-60.416$ to $137.220$~s. 
Our results are consistent with those by~\citet{Chen21}.
Our analysis shows a hard-to-soft evolution, modulated by $f/f_{\rm tot}$ tracking the light curve peaks (Figure~\ref{fig:180720B}).
As a consequence, the fluence of the brightest peaks is underestimated by up to 40\%, whereas the fluence of the weakest ones is overestimated by a comparable amount.
Among the examined GRBs, 180720B, 190114C, and 171010A exhibited the larger deviations of $f/f_{\rm tot}$ from 1, reaching 60-70\%, indicating that the effect is probably stronger for brightest GRBs. We can therefore safely assume that neglecting the spectral evolution for the bulk of GRBs considered in this work has a milder effect than what we assessed for the brightest cases.

\begin{figure}
 \includegraphics[width=1\columnwidth]{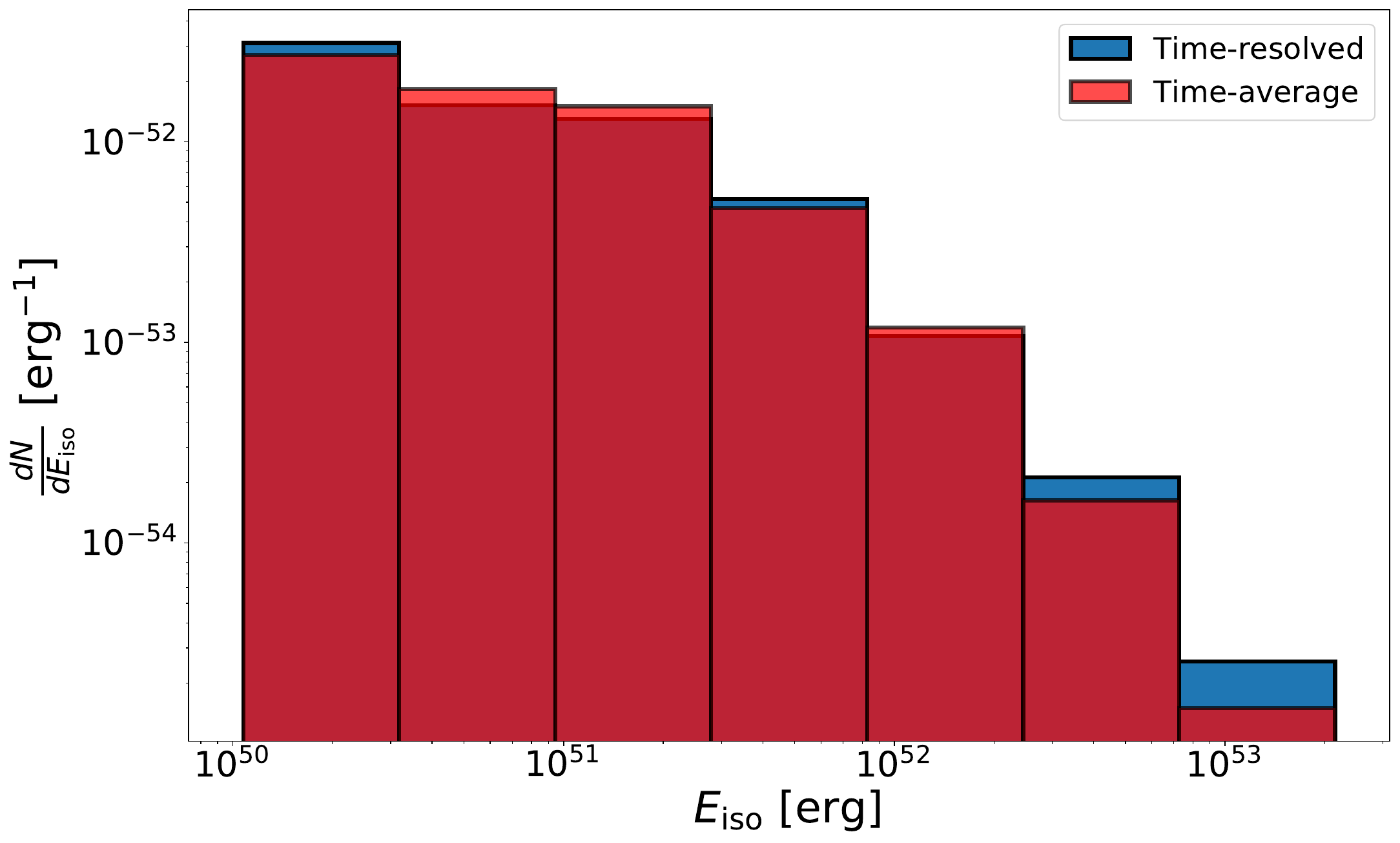}
 \caption{$E_{\rm{iso}}$ distributions obtained with either time-resolved (blue) or time-average (red) spectral analysis.}
\label{fig:timeres_vs_timeave_dist}
\end{figure}

 We finally explored to what extent the $E_{\rm{iso}}$ distribution was affected by ignoring the spectral evolution: we compared $E_{\rm iso}$ calculated with both time-resolved and time-average analysis and derived the corresponding distributions (Figure~\ref{fig:timeres_vs_timeave_dist}). We ran statistical tests to assess the null hypothesis that the two $E_{\rm{iso}}$ distributions are drawn from the same population. A KS test yielded a p-value of $0.87$, while the AD test gave a lower-limit on the p-value $> 0.25$. We concluded that we cannot reject the hypothesis that the two distributions are drawn from the parent population, so we can safely ignore the bias introduced by neglecting the spectral evolution.

Additionally, we compared the best-fit values for the PL index $\alpha$ that models the tail of each distribution: we obtained $\alpha_{\rm tr} = 1.62^{+0.20}_{-~0.18}$ and other $\alpha_{\rm ta} = 1.63^{+~0.19}_{-0.19}$ for the time-resolved and for the time-average distribution, respectively. The two values are indistinguishable within uncertainties, which confirm that neglecting the no-spectral evolution assumption is safe.

\bibliography{alles_grbs}{}
\bibliographystyle{aasjournal}

\end{document}

%% file: autors.tex
\author[0000-0002-8799-2510]{R.~Maccary}
\affiliation{Department of Physics and Earth Science, 
University of Ferrara, Via Saragat 1, I-44122 Ferrara, Italy}

\author[0000-0001-6869-0835]{C.~Guidorzi}
\affiliation{Department of Physics and Earth Science, University of Ferrara, Via Saragat 1, I-44122 Ferrara, Italy}
\affiliation{INFN -- Sezione di Ferrara, Via Saragat 1, 44122 Ferrara, Italy}
\affiliation{INAF -- Osservatorio di Astrofisica e Scienza dello Spazio di Bologna, Via Piero Gobetti 101, 40129 Bologna, Italy}

\author[0000-0001-5355-7388]{L.~Amati}
\affiliation{INAF -- Osservatorio di Astrofisica e Scienza dello Spazio di Bologna, Via Piero Gobetti 101, 40129 Bologna, Italy}

\author[0000-0003-0727-0137]{L.~Bazzanini}
\affiliation{Department of Physics and Earth Science,  University of Ferrara, Via Saragat 1, I-44122 Ferrara, Italy}
\affiliation{INAF -- Osservatorio di Astrofisica e Scienza dello Spazio di Bologna, Via Piero Gobetti 101, 40129 Bologna, Italy}

\author[0000-0002-8255-5127]{M.~Bulla}
\affiliation{Department of Physics and Earth Science, University of Ferrara, Via Saragat 1, I-44122 Ferrara, Italy}
\affiliation{INFN -- Sezione di Ferrara, Via Saragat 1, 44122 Ferrara, Italy}
\affiliation{INAF, Osservatorio Astronomico d’Abruzzo, via Mentore Maggini snc, 64100 Teramo, Italy}

\author[0000-0002-4200-1947]{A.~E.~Camisasca}
\affiliation{Department of Physics and Earth Science, 
University of Ferrara, Via Saragat 1, I-44122 Ferrara, Italy}

\author[0009-0006-1140-6913]{L.~Ferro}
\affiliation{Department of Physics and Earth Science, University of Ferrara, Via Saragat 1, I-44122 Ferrara, Italy}
\affiliation{INAF -- Osservatorio di Astrofisica e Scienza dello Spazio di Bologna, Via Piero Gobetti 101, 40129 Bologna, Italy}

\author[0000-0003-2284-571X]{F.~Frontera}
\affiliation{Department of Physics and Earth Science, University of Ferrara, Via Saragat 1, I-44122 Ferrara, Italy}
\affiliation{INAF -- Osservatorio di Astrofisica e Scienza dello Spazio di Bologna, Via Piero Gobetti 101, 40129 Bologna, Italy}

\author[0000-0003-0292-6221]{A.~Tsvetkova}
\affiliation{Department of Physics, University of Cagliari, SP Monserrato-Sestu, km 0.7, 09042 Monserrato, Italy}
\affiliation{INAF -- Osservatorio di Astrofisica e Scienza dello Spazio di Bologna, Via Piero Gobetti 101, 40129 Bologna, Italy}
%\affiliation{Ioffe Institute, Politekhnicheskaya 26, 194021 St. Petersburg, Russia}